\documentclass[]{pasj02}
\usepackage[switch,mathlines]{lineno} 
\usepackage{natbib}
\usepackage{url}
\usepackage{subcaption}

\newcommand{\psfupdate}[1]{{#1}}

\jyear{2025}
\Received{}
\Accepted{}

\begin{document}

\title{Pre-flare and active region plasma flows and structure seen by the short wavelength camera on SOLAR-C/EUVST}

\author{
    James \textsc{McKevitt},\altaffilmark{1,2}\altemailmark\orcid{0000-0002-4071-5727}\email{james.mckevitt.21@ucl.ac.uk}
    Sarah \textsc{Matthews},\altaffilmark{1}\orcid{0000-0001-9346-8179}
    David H. \textsc{Brooks},\altaffilmark{3,1}\orcid{0000-0002-2189-9313}
    Toshifumi \textsc{Shimizu},\altaffilmark{4}\orcid{0000-0003-4764-6856}
    Akiko \textsc{Tei},\altaffilmark{5}\orcid{0000-0002-0236-361X}
    Ignacio \textsc{Ugarte-Urra},\altaffilmark{6}\orcid{0000-0001-5503-0491}
    Shinsuke \textsc{Imada},\altaffilmark{7}\orcid{0000-0001-7891-3916}
    Shin \textsc{Toriumi},\altaffilmark{4}\orcid{0000-0002-1276-2403}
    Charles M. \textsc{Brown},\altaffilmark{6}
    Ryohko \textsc{Ishikawa},\altaffilmark{5}\orcid{0000-0001-8830-0769}
    Yukio \textsc{Katsukawa},\altaffilmark{5}\orcid{0000-0002-5054-8782}
    Hirohisa \textsc{Hara},\altaffilmark{5}\orcid{0000-0001-5686-3081}
    Duncan \textsc{Rust},\altaffilmark{1}
    David \textsc{Walton},\altaffilmark{1}
    Berend \textsc{Winter},\altaffilmark{1}
    Deborah \textsc{Baker},\altaffilmark{1}\orcid{0000-0002-0665-2355}
    Hamish \textsc{Reid},\altaffilmark{1}\orcid{0000-0002-6287-3494}
    Peter \textsc{Young},\altaffilmark{8}\orcid{0000-0001-9034-2925}
    Tiago M. D. \textsc{Pereira},\altaffilmark{9,10}\orcid{0000-0003-4747-4329}
    Louisa \textsc{Bradley},\altaffilmark{1}
    Alexey \textsc{Shitvov},\altaffilmark{1}
    Louise \textsc{Harra},\altaffilmark{11,12}\orcid{0000-0001-9457-6200}
    International SOLAR-C team
}

\altaffiltext{1}{University College London, Mullard Space Science Laboratory Holmbury St Mary, Dorking Surrey, RH5 6NT, UK}
\altaffiltext{2}{University of Vienna, Institute of Astrophysics Türkenschanzstrasse 17 Vienna A-1180, Austria}
\altaffiltext{3}{Computational Physics, Inc., Springfield, VA 22151, USA}
\altaffiltext{4}{Institute of Space and Astronautical Science, Japan Aerospace Exploration Agency, 3-1-1 Yoshinodai, Chuo-ku, Sagamihara, Kanagawa 229-8510, Japan}
\altaffiltext{5}{National Astronomical Observatory of Japan, Mitaka, Tokyo 181-8588, Japan}
\altaffiltext{6}{Space Science Division, Naval Research Laboratory, Washington, DC 20375, USA}
\altaffiltext{7}{Institute for Space-Earth Environmental Research, Nagoya University, Furo-cho, Chikusa-ku, Nagoya 464-8601, Japan}
\altaffiltext{8}{NASA Goddard Space Flight Center, Solar Physics Laboratory, Heliophysics Science Division, Greenbelt, MD 20771, USA}
\altaffiltext{9}{Rosseland Centre for Solar Physics, University of Oslo, Oslo, Postboks 1029, 0315, Norway}
\altaffiltext{10}{Institute of Theoretical Astrophysics, University of Oslo, PO Box 1029, Blindern 0315, Oslo, Norway}
\altaffiltext{11}{Physikalisch-Meteorologisches Observatorium Davos/World Radiation Center, PMOD/WRC, Dorfstrasse 33, Davos Dorf, 7260, GR, Switzerland}
\altaffiltext{12}{D-PHYS, ETH Zürich, Wolfgang-Pauli Strasse 27, Zürich, 8093, ZH, Switzerland}

\KeyWords{Sun: corona --- Sun: flares --- Sun: UV radiation --- instrumentation: spectrographs --- space vehicles: instruments}

\maketitle

\begin{abstract}
The mechanisms triggering solar flares and driving coronal heating occur across wide temperature ranges on small spatial scales and short timescales, making them difficult to observe with current instrumentation. The upcoming SOLAR-C mission, launching in the late 2020s, will provide unprecedented plasma diagnostic capability with its high-throughput extreme-ultraviolet (EUV) spectroscopic telescope (EUVST), capable of $\sim$0.2~arcsec~pix$^{-1}$ spatial sampling ($\sim$0.4~arcsec resolution), continuous temperature coverage from 0.02--15~MK, and exposure times down to 0.5~seconds. We present forward modelling of the spectrograph's short wavelength camera (170--210~\AA{}; SOLAR-C/EUVST-SW) and its response to log~T$\sim$6.2 coronal plasma in a three-dimensional MHD-simulated pre-flare active region. We compare this performance to that of the previous-generation EUV Imaging Spectrometer (EIS) on Hinode (SOLAR-B). Our results demonstrate that SOLAR-C/EUVST can distinguish individual flux tubes in simulated active region loops which Hinode/EIS cannot resolve. In simulated pre-flare plasma, SOLAR-C/EUVST captures sharp velocity gradients between adjacent upflowing and downflowing plasma which Hinode/EIS is unable to resolve. Doppler velocity measurement accuracy will reach better than 1~km~s$^{-1}$ 
in active regions. We show that this next-generation spectrograph can be expected to directly observe processes potentially related to flare triggering, such as plasma flows from low-altitude reconnection linked to emerging flux, and determine whether active region loops consist of a small number of strands or the hundreds predicted by magnetic reconnection-induced nanoflare heating models.
\end{abstract}


\section{Introduction}\label{sec:introduction}

The structure of the solar atmosphere \citep{shapiro_suns_2019}, the build up and explosive release of energy in the corona \citep{mckevitt_link_2024}, and the driving of mid- to upper-chromospheric and coronal plasma by the photospheric and lower-chromospheric magnetic field \citep{wiegelmann_solar_2021}, are all aspects of the Sun which have been well observed and well studied. The driving mechanisms behind these phenomena, however, happen rapidly on small spatial scales and across a wide temperature range, making their detailed characterisation challenging. While the launch of Hinode \citep[SOLAR-B;][]{kosugi_hinode_2007} in 2006 marked a revolution in our ability to understand coronal plasma \citep{hinode_review_team_achievements_2019}, further improvements in instrumentation are needed. In the late 2020s the SOLAR-C spacecraft will fly a next-generation extreme ultraviolet (EUV) spectrometer, capable of resolving very small spatial scales (300~km at the Sun; $\sim$0.4~arcsec) coherently and simultaneously across a wide temperature range (0.02~MK--15~MK) and on very short timescales (exposure times down to 0.5~seconds). Such performance will allow us to directly observe the fundamental processes taking place in the chromosphere and corona, resolving the key outstanding questions in solar physics, chief among them concerning the heating of the corona and the onset of solar flares.


\subsection{Coronal heating}

It is thought that the heating of the corona to puzzlingly high temperatures is impulsive, where such events regardless of their mechanism can be termed nanoflares \citep{klimchuk_key_2015}. There are two primary mechanisms thought to be behind these events, namely magnetic flux tube braiding caused by random photospheric motions driving reconnection in the corona, and MHD wave heating, whereby MHD waves such as Alfvén waves are damped into heat \citep[e.g.,][]{klimchuk_key_2015}.

In the first scenario, different adjacent magnetic flux tubes, each carrying relatively homogeneous plasma along the field, reconnect with one another on small scales, causing impulsive and localised heating \citep{parker_nanoflares_1988}. While individual events are stochastic and impulsive, their intermediate-to-high frequency across many flux tubes produces the appearance of quasi-continuous heating, giving rise to the bright and relatively steady emission of the diffuse corona. By contrast, low-frequency heating events, in which plasma cools fully before the next event, may dominate within distinct coronal loops, where they produce the strongly time-variable brightenings commonly observed \citep{klimchuk_key_2015}. Since it was first proposed, computational advances have demonstrated that magnetic field braiding-induced heating is a viable mechanism for coronal heating \citep[e.g.,][]{mondal_spatial_2025}, and improving observational capabilities continue to provide mounting evidence \citep[e.g.,][]{ishikawa_detection_2017}. However, it has not yet been conclusively established that such heating is both ubiquitous throughout the solar corona and sufficient to sustain its temperature.

In the second scenario, MHD waves propagate energy injected in the low solar atmosphere up and into the corona, which is then released as the waves are damped \citep{van_doorsselaere_coronal_2020}. Several dissipation mechanisms have been proposed, such as resonant absorption, Alfvén wave turbulence, and Kelvin-Helmholtz instabilities (see \cite{van_doorsselaere_coronal_2020} for details), and MHD wave heating models have been found sufficient to explain quiet sun heating \citep[e.g.,][]{shi_first_2021}. However, MHD wave heating models are not (yet) sufficient to explain active region coronal heating alone, and require improved observations to inform theory and mature further.

The observational support for nanoflares so far comes from signs of strands of possible field braiding \citep{cirtain_energy_2013}, intermittent single-temperature small-scale brightening at loop footpoints indicative of nanoflares higher up in the corona \citep{testa_observing_2013}, and small-scale high-temperature thermal emission seen in a quiescent active region indicative of impulsive nanoflare heating \citep{ishikawa_detection_2017}. These studies required the use of short sounding rocket flights, which are the only instruments to have so far achieved sufficient spatial resolution and thermal range, with the High-Resolution Coronal Imager \citep[Hi-C;][]{kobayashi_high-resolution_2014} providing $\sim$0.1~arcsec/pixel and the Focusing Optics X-ray Solar Imager \citep[FOXSI;][]{christe_foxsi-2_2016} coverage up to $\sim$20~MK for their respective studies. More recently, space-based observations have identified nanojets supporting reconnection-driven heating \citep{antolin_reconnection_2021, sukarmadji_observations_2022}, but their observation is very challenging with the spatial and temporal resolution of current space-based instrumentation. In the case of MHD wave heating, modelling is relatively advanced \citep[e.g.,][]{antolin_fine_2014} and there are some supporting observations available \citep{okamoto_resonant_2015,antolin_resonant_2015}. Better supporting observations would require continual tracking of waves through the atmosphere at small scales, and while the space-based Interface Region Imaging Spectrograph \citep[IRIS;][]{de_pontieu_interface_2014} is capable of resolving $\sim$0.4~arcsec threads in the interface region for long durations, only sounding rocket-based telescopes like Hi-C can currently track such small features at such a high resolution, at coronal temperatures. However, its flights are on the order of $\sim$200~seconds, insufficient to properly observe any waves and their dissipation.

\subsection{Plasma instabilities (flares)}


Both flux tube braiding and MHD wave heating mechanisms are driven by the photosphere injecting Poynting flux up into the atmosphere above. The $\beta\ll1$ plasma in the corona is confined to the magnetic field structures rooted in the $\beta\gg1$ plasma below in the photosphere and lower chromosphere, where $\beta$ is the ratio of the gas pressure to magnetic pressure. Under continued stressing by large-scale motions in the $\beta\gg1$ plasma the coronal magnetic field is driven away from its lowest-energy (potential) state. In this way, free magnetic energy gradually accumulates in the corona and is stored in non-potential field structures, from which the energy can later be released explosively during flares and coronal mass ejections \citep{mckevitt_link_2024}.

The plasma tracing coronal field lines appears to hold clues about when this large-scale energy release will happen. For example, small-scale brightenings beginning before the main onset of a flare and interpreted as being related to flare triggering have been seen previously \citep[e.g.,][]{warren_ultraviolet_2001}. There are also several key spectroscopic characteristics of coronal plasma that have been found to precede a solar flare \citep{harra_coronal_2023}. Non-thermal broadening, that being the excess broadening of spectral lines beyond that explained by instrumental and thermal effects, is seen to increase before a solar flare \citep[e.g.,][]{harra_coronal_2009, harra_location_2013, toriumi_flare-productive_2019}. It has also been shown that plasma begins upflowing before a flare \citep[e.g.,][]{imada_coronal_2014,bamba_study_2017,woods_observations_2017}.

While there is certainly some connection between these spectroscopic signatures and the triggering of plasma instabilities, current limits primarily in spatial and thermal resolution mean that understanding the physical mechanisms behind them is difficult. Non-thermal line broadening, for example, appears somewhat independent of spatial resolution down to the levels of our best current instrumentation $\sim$0.3~arcsec \citep[e.g.,][]{de_pontieu_why_2015, testa_high_2016}, and is thought to encompass different physical processes such as unresolved Alfvénic motions \citep{banerjee_signatures_2009}, turbulence induced by impulsive heating events \citep{patsourakos_nonthermal_2006}, and the superposition of unresolved small-scale flows \citep{doschek_flows_2008}, amongst others. When seen before a flare, it is thought early-onset reconnection triggers turbulence causing the broadening \citep{joshi_pre-flare_2011}, as is similarly thought for pre-flare brightenings \citep{woods_observations_2017}. Recent work by \cite{russell_solar_2025} suggests ion temperatures during flares could be much higher than previously thought, contributing to broadening more than previously expected. There is also a known connection between the observed broadening and upflowing plasma \citep[e.g.,][]{doschek_dynamics_2012}, which is widely accepted to be at least partly caused by unresolved upflows at high velocities \citep{doschek_flows_2008, del_zanna_solar_2018}. The analysis is further complicated by the differing spatial resolutions of chromospheric and coronal observations, since separate instruments are required to span the full temperature range (IRIS and Hinode/EIS respectively). A high and consistent spatial resolution is required to fully understand the pre-flare mechanisms behind these signatures.

At the point of the flare onset, in the case of eruptive flares, a standard model for such eruptions envisions reconnection occurring above flare arcades causing hot fast plasma outflows \citep{shibata_hot-plasma_1995}. \cite{imada_evidence_2013} found evidence of $\sim$30~MK plasma flowing at over 500~km/s above flare loops consistent with this model using Hinode/EIS, and found such structure to have notable extent and complexity which could not be fully resolved in space or temperature with Hinode/EIS. They also noted the weakness of the signal given the reconnection region appeared to be where plasma emission was relatively low, meaning a higher throughput telescope is required to observe such reconnection regions directly.

Computational modelling has so far been primarily concentrated on connecting quiescent active region heating physics with spectroscopic signatures \citep[e.g.,][]{pontin_non-thermal_2020,asgari-targhi_observations_2024}, and no direct links between these (pre-)flare mechanisms and the forward-modelled spectroscopic signatures have yet been made. However, magnetohydrodynamic (MHD) modelling of solar flares has improved in recent years \citep[e.g.,][]{cheung_comprehensive_2018}, making it possible to start disentangling the signatures we see with current instrumentation and hunt for new signatures that the next generation of spectrographs will see.


\subsection{SOLAR-C}\label{sec:intro_solarc}

To understand the mechanisms powering coronal heating, to understand pre-flare mechanisms related to the triggering of solar flares, and to fully resolve the result of magnetic reconnection on the solar atmosphere, a new generation of spectrometer is needed.

Work first began around the mid to late 2000s with the formation of the SOLAR-C working group to design the successor spacecraft to Hinode \citep{solar-c_working_group_interim_2011}\footnote{\url{https://hinode.nao.ac.jp/SOLAR-C/SOLAR-C/Documents/Interim2011/SC_Interim_all.pdf}}. The design converged on by the group centred around a high-throughput telescope capable of observing a wide temperature range of plasma simultaneously with minimal gaps at a high spatial and temporal resolution \citep{teriaca_lemur_2012}. The Next Generation Solar Physics Mission Science Objectives Team (NGSPM-SOT) was then formed in 2016 between United States National Aeronautics and Space Administration (NASA), Japan Aerospace Exploration Agency (JAXA), and European Space Agency (ESA) to study and report on a multilateral solar physics mission concept \citep{ngspm-sot_ngspm-sot_2017}\footnote{\url{https://hinode.nao.ac.jp/SOLAR-C/SOLAR-C/Documents/NGSPM_report_170731.pdf}}. Based on this work, SOLAR-C/EUVST (EUV high-throughput Spectroscopic Telescope) is now under construction and nearing its launch in the late-2020s \citep{shimizu_solar-c_2020}, and is set to deliver coverage of plasma with a wide temperature range (0.02--15~MK) with small gaps ($\Delta$log~T$\leq$0.4) at a high spatial resolution (down to $\sim$0.4~arcsec; $\sim$300~km at the Sun) and temporal sampling rate (0.5~s at its fastest). The SOLAR-C spacecraft bus and telescope development is led by ISAS/JAXA in Japan with the EUVST spectrograph's short-wavelength camera (SW; 170~\AA{}--210\AA{}) provided by ESA, the long-wavelength camera (LW; 690~\AA{}--1275~\AA{}) by NASA, and additional instrumentation and components provided by NASA and individual European countries.


The scientific objectives of SOLAR-C are centred on two themes \citep{shimizu_solar-c_2020}: to I) understand how fundamental processes lead to the formation of the solar atmosphere and solar wind, and to II) understand how the solar atmosphere becomes unstable, releasing the energy that drives solar flares and eruptions.

\subsection{This paper}

SOLAR-C/EUVST will be one of the most powerful plasma diagnostic tools ever used by solar physicists once it is launched, and so some preparatory work is required to understand the data it will produce. 

We present the expected performance of the short wavelength camera (SOLAR-C/EUVST-SW), being built for ESA at UCL's Mullard Space Science Laboratory (MSSL), and its ability to resolve plasma in space. We do this by synthesising Fe~XII~log~T$\sim$6.2 coronal plasma emission from a simulated 3D MHD atmosphere, then simulating the instrument response and examining the structures and flows that we expect to see in quiescent active region structures and in a pre-flaring part of an active region. We reserve an analysis of the spectrograph performance in time and temperature for future work including the aforementioned long-wavelength channel. We make available the complete forward-modelling \texttt{ECLIPSE} code (Emission Calculation and Line Prediction for SOLAR-C EUVST) and a Python API with tutorials and guides so other pre-launch science can be done by the community\footnote{This paper uses version \psfupdate{0.6.1.4: \url{https://doi.org/10.5281/zenodo.19662222}}. The source code and ongoing development are available at: \url{https://github.com/jamesmckevitt/eclipse}}.


\section{Method}\label{sec:method}

In this section we present the three-dimensional solar atmosphere we use in this paper (Section~\ref{sec:method_mhdsim}), the technique we use to synthesise the EUV emission of that atmosphere (Section~\ref{sec:method_synthesis}), and the details of the method used to simulate the response of SOLAR-C/EUVST to the EUV emission (Section~\ref{sec:method_response}).

\subsection{Simulated Solar Atmosphere}\label{sec:method_mhdsim}

We use publicly-available snapshots from the three-dimensional solar atmosphere of \cite{cheung_comprehensive_2018}, simulated using the 3-D radiative magnetohydrodynamic (MHD) code MURaM.

MURaM \citep[Max Planck Institute for Solar System Research/University of Chicago Radiation MHD;][]{vogler_simulations_2005} was originally developed to model how magnetic fields interact with convection in the photosphere. It solves the full set of compressible magnetohydrodynamic equations on a uniform Cartesian grid with fourth-order accuracy in space and time. The code computes radiative transfer along many short rays through the grid using wavelength-dependent opacities, and includes partial ionisation using a tabulated equation of state. It uses periodic side boundaries and a lower boundary that injects heat, while the top boundary is closed but allows magnetic field lines to remain vertical. \cite{rempel_extension_2017} extended MURaM beyond the photosphere and into the million-degree solar corona. Energy losses at high temperatures are treated using the CHIANTI atomic database \citep{landi_chiantiatomic_2012}. Heat is allowed to flow along magnetic field lines using Spitzer conductivity, and the \lq{}Boris correction\rq{} is used to limit the Alfvén speed to prevent the numerical timesteps becoming prohibitively small \citep{boris_physically_1970}. Classical conduction is rewritten so that heat propagates at a finite speed as in \cite{snodin_simulating_2006}.

\cite{cheung_comprehensive_2018} simulated several hours of solar evolution using the MURaM code. They presented a simulation domain approximately 100~Mm in $x$, 50~Mm in $y$, and in $z$ from approximately 8~Mm below to 42~Mm above the photosphere, with a grid spacing of 192~km in the horizontal directions and 64~km vertically, capturing both the convective driving and overlying atmosphere in one continuous box. The magnetic field is set up to provide the key ingredients for a flare-productive active region. The initial condition is a relaxed bipolar sunspot pair with 3.4$\times$10$^{21}$~Mx per polarity. A second, strongly twisted bipole carrying 9.5$\times$10$^{20}$~Mx of horizontal flux (approximately 30\% of the pre-existing active region flux; the associated vertical flux is self-consistently computed) is then emerged through the lower boundary by imposing a 200~m/s upflow within an ellipsoidal patch. This is designed to resemble the flare-productive parasitic flux emergence observed north of the leading spot in active region NOAA~12017 (see \citealt{chintzoglou_origin_2019}). After the emergence begins, convection and Lorentz forces shear and stress the field until a coronal flux rope forms above the parasitic polarity inversion line, eventually destabilising and driving a flare-like energy release. The resulting atmosphere contains dense chromospheric plasma, compact flare loops reaching $>$100~MK, long-range coronal connections (made possible by the periodic side boundaries), and regions of intensity dimming produced by strong heating along reconnected field lines. In this way, the model evolves from a quiet active region into a pre-eruption configuration and then through a full eruptive event, producing a stratified, magnetically complex atmosphere that closely resembles a real solar flare.

In this paper we consider the instrument response to the atmosphere after the emergence of the parasitic bipole and approximately 30~minutes before the peak in simulated GOES X-ray emission. At this time the atmosphere contains quiescent active region plasma, but also a complex magnetic topology and enhanced coronal heating above the newly-emerged bipole in the pre-flare atmosphere. This allows us to consider the performance of SOLAR-C/EUVST when observing typical active region conditions, and its ability to resolve signatures preceding a solar flare.

\subsection{Synthesis of Spectral Line Intensities}\label{sec:method_synthesis}

The results of the MHD simulations described above provide self-consistent values for the temperature, density, and velocity of the solar plasma in the pre-flare active region. Our instruments observe this plasma using remote sensing of its thermal emission, and so we first need to synthesise such emission. In this study we are concerned with optically-thin log~T$\sim$6.2 plasma, and so perform an optically-thin line synthesis.

\subsubsection{Definition of Line Emission}

The emissivity of a specific atomic transition from an upper atomic level $j$ to a lower atomic level $i$ is given fundamentally by the product of the number density of ions occupying the excited state, $N_j(X^{+m})$, and the Einstein spontaneous emission coefficient, $A_{ij}$:

\begin{equation}
    \varepsilon_{ij} = N_j(X^{+m})\, A_{ij} \quad [\text{ph s}^{-1}\text{ cm}^{-3}].
\end{equation}

Integrating this emissivity along the line-of-sight (LOS) direction $h$, the intensity emitted per steradian is

\begin{equation}
    I_{ij} = \frac{1}{4\pi} \int_h N_j(X^{+m}) A_{ij}\,dh \quad [\text{ph s}^{-1}\text{ cm}^{-2}\text{ sr}^{-1}],
\end{equation}

\noindent{}where we can multiply by the photon energy $h\nu$ to convert this photon intensity to an energy intensity

\begin{equation}
    I_{ij} = \frac{h\nu}{4\pi} \int_h N_j(X^{+m}) A_{ij}\,dh \quad [\text{erg s}^{-1}\text{ cm}^{-2}\text{ sr}^{-1}].
\end{equation}

\subsubsection{Contribution Functions}

It is convenient to consider the excited-level ion number density \(N_j(X^{+m})\) relative to the ion number density \(N(X^{+m})\), the elemental number density \(N(X)\), the hydrogen number density \(N(H)\), and the electron number density \(N_e\), in the following form:

\begin{equation}
    N_j(X^{+m}) = \frac{N_j(X^{+m})}{N(X^{+m})} \frac{N(X^{+m})}{N(X)} \frac{N(X)}{N(H)} \frac{N(H)}{N_e} N_e \quad [\text{cm}^{-3}],
\end{equation}

\noindent{}as when we consider the contribution function defined as 

\begin{multline}
    G_{ij}(T,N_e)= h\nu\,\frac{N_j(X^{+m})}{N(X^{+m})}\frac{N(X^{+m})}{N(X)}\frac{N(X)}{N(H)}\frac{N(H)}{N_e^{2}}\,A_{ij} \\
    [\mathrm{erg\,cm^{3}\,s^{-1}}],
\end{multline}

\noindent{}we can simplify our expression for the intensity of emitted radiation to

\begin{equation}
    I_{ij} = \frac{1}{4\pi}\int_h G_{ij}(T,N_e){N_e}^2\, dh \quad [\text{erg s}^{-1}\text{ cm}^{-2}\text{ sr}^{-1}].
\end{equation}

The contribution function \(G_{ij}(T, N_e)\) combines atomic physics, ionisation equilibrium, excitation processes, elemental abundances, and radiative transitions into a single factor, conveniently tabulated by atomic databases such as CHIANTI. For our work here, we use CHIANTI version 10.0 \citep{del_zanna_chiantiatomic_2021} \psfupdate{through the fiasco Python interface \citep{will_barnes_fiasco_2025}}.





\subsubsection{Differential Emission Measure (DEM)}


In practice, the integration along the LOS is numerically evaluated using discrete simulation voxels (3D volume elements). To account for this, we introduce the Differential Emission Measure (DEM) which quantifies how much emitting material exists at each temperature along the LOS:

\begin{equation}
    \xi(T) = {N_e}^2\frac{dh}{dT} \quad [\text{cm}^{-5}\,\text{K}^{-1}].
\end{equation}

This transforms our integral along the spatial coordinate $h$ into an integral over temperature $T$:

\begin{equation}
    I_{ij} = \frac{1}{4\pi}\int_T G_{ij}(T,\langle N_e\rangle_T)\,\xi(T)\, dT \quad [\text{erg s}^{-1}\text{ cm}^{-2}\text{ sr}^{-1}],
\end{equation}

\noindent{}where we explicitly use the emission-measure-weighted mean electron density $\langle N_e\rangle_T$ for consistency with DEM calculations.

Numerically, we compute this DEM by binning the simulation data into discrete temperature intervals $(T_i,T_{i+1})$ using a Riemann summation \citep{riemann_ueber_1867}, summing the squared electron densities along the LOS axis:

\begin{align}
    \xi(T_i) =\ & \frac{1}{\Delta\log_{10}T} \sum_{z} {N_e(x,y,z)}^2 \Delta h \notag \\
    &\times \Theta(T_i \le T(x,y,z) < T_{i+1})\quad[\text{cm}^{-5}\text{ dex}^{-1}],
\end{align}

\noindent{}with the Heaviside step function $\Theta$ \citep{heaviside_electrical_1892} selecting voxels within each temperature bin. Because we bin the plasma in logarithmic temperature, $\xi(T_i)$ is defined per $\log_{10}T$ bin and therefore has units $\text{cm}^{-5}\text{ dex}^{-1}$, where $\text{dex}$ denotes a bin of width $\Delta\log_{10}T$.

\subsubsection{Two-dimensional DEM and Velocity Distribution}\label{sec:2d_dem_equation}

To synthesise realistic line profiles, we must incorporate the plasma velocity distribution along the LOS. Thus, we extend the DEM into two-dimensional temperature-velocity space (see \citealt{newton_testing_1996} for an observational approach), defining:

\begin{equation}
    \Xi(T,v) = {N_e}^2\frac{\partial^2 h}{\partial T \partial v} \quad [\text{cm}^{-6}\,\text{K}^{-1}\,\text{s}],
\end{equation}

\noindent{}where $v$ is the line-of-sight velocity component. Numerically, this is computed using a two-dimensional histogram, simultaneously binning the voxelised temperature and velocity data:

\begin{multline}
    \Xi(T_i,v_j) = \sum_z {N_e(x,y,z)}^2\Delta h\,\Theta(T_i\le T(x,y,z)<T_{i+1}) \times \\
    \Theta(v_j\le v(x,y,z)<v_{j+1}) \quad [\text{cm}^{-5}],
\end{multline}

\noindent{}implemented efficiently using Einstein summation notation \citep{einstein_grundlage_1916} in the numerical code via:

\begin{equation}
    \Xi_{ijlm} = \sum_k ({N_e}^2\Delta h)_{ijk}\,\Theta_{ijkl}^{(T)}\,\Theta_{ijkm}^{(v)} \quad [\text{cm}^{-5}],
\end{equation}

\noindent{}where indices $i,j$ here represent spatial pixels horizontal to the observing plane $(x,y)$, $k$ the LOS index $(z)$, and $l,m$ temperature and velocity bins respectively. In our implementation we parallelise this using Dask\footnote{\url{http://dask.pydata.org}} \citep{rocklin_dask_2015}.

Because each element $\Xi(T_i,v_j)$ already integrates over its full bin, no further bin-width factors (such as $\Delta T_i$ or $\Delta v_j$) are required in subsequent numerical integrations.

\subsubsection{Thermal Broadening and Line Profiles}

The thermal Doppler broadening of emission lines arises from the Maxwell-Boltzmann velocity distribution \citep{maxwell_v_1860,maxwell_ii_1860,boltzmann_weitere_1872}. The thermal width $\Delta\lambda_D$ is given by

\psfupdate{
\begin{equation}
    \Delta\lambda_D = \frac{\lambda_0}{c}\sqrt{\frac{k_B T}{m_{\text{ion}}}}\quad[\text{cm}],
\end{equation}
}

\noindent{}where $k_B$ is Boltzmann's constant and $m_{\text{ion}}$ the mass of the emitting ion.

The Gaussian line profile, normalised such that $\int\phi(\lambda,T,v)d\lambda=1$, is then:

\begin{equation}
    \phi(\lambda,T,v) = \frac{1}{\sqrt{2\pi}\Delta\lambda_D}\exp\!\left[-\frac{(\lambda - \lambda_0(1+v/c))^2}{2\Delta\lambda_D^2}\right]\quad[\text{cm}^{-1}].
    \label{equ:synthetic_gaussian_line_profile}
\end{equation}

\subsubsection{Final Intensity Integration}

Combining all the above, the final expression for the specific intensity as a continuous function of wavelength $\lambda$ is

\begin{multline}
    I_{ij}(\lambda) = \frac{1}{4\pi}\iint G_{ij}(T,\langle N_e\rangle_T)\,\Xi(T,v)\,\phi(\lambda,T,v)\,dT\,dv \\
    [\text{erg s}^{-1}\text{ cm}^{-2}\text{ sr}^{-1}\text{ cm}^{-1}],
\end{multline}

\noindent{}discreetly numerically approximated as

\begin{multline}\label{equ:final_synth_intensity}
    I_{ij}(\lambda) = \frac{1}{4\pi}\sum_{i,j} G_{ij}(T_i,\langle N_e\rangle_{T_i})\,\Xi(T_i,v_j)\,\phi(\lambda,T_i,v_j) \\
    [\text{erg s}^{-1}\text{ cm}^{-2}\text{ sr}^{-1}\text{ cm}^{-1}].
\end{multline}

The integrated line intensity can be calculated by simply considering the intensity across the whole wavelength range where the integrated line intensity

\begin{equation}
  \mathcal{I}_{ij}=\int I_{ij}(\lambda)\,d\lambda  \quad [\text{erg s}^{-1}\text{ cm}^{-2}\text{ sr}^{-1}],
\end{equation}

\noindent{}and is numerically approximated as

\begin{equation}
  \mathcal{I}_{ij}=\sum_{k=1}^{N_\lambda} I_{ij}(\lambda_k)\,\Delta\lambda_k  \quad [\text{erg s}^{-1}\text{ cm}^{-2}\text{ sr}^{-1}].
\end{equation}

\begin{figure*}
    \centering
    \includegraphics[width=\linewidth]{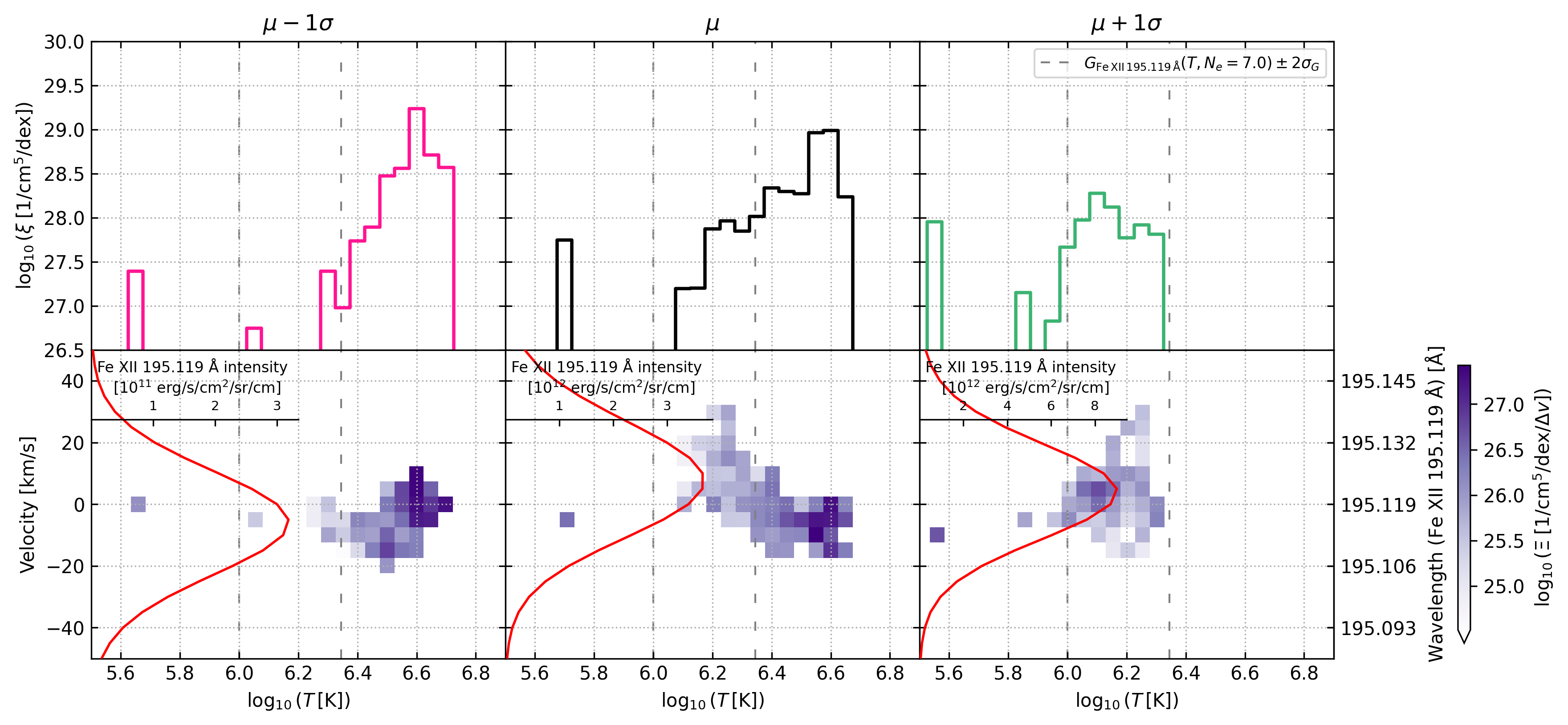}
    \caption{Differential emission measure ($\xi$; top row) and two-dimensional emission measure ($\Xi$; bottom row) distributions for three representative pixels in the simulation: mean minus one standard deviation ($\mu-1\sigma$, left), mean ($\mu$, center), and mean plus one standard deviation ($\mu+1\sigma$, right) of the total Fe~XII~195.119~\AA{}, FeXII~195.179~\AA{}, and background line intensities. Vertical dashed lines indicate the temperature range contributing $\sim$95\% of the Fe~XII~195.119~\AA\ line intensity, from the CHIANTI contribution function. The corresponding synthetic Fe~XII~195.119~\AA\ line profiles are overplotted on the two-dimensional emission measures in red, corresponding to the bottom row's y axes and the inset x axis.\\
    Alt text: Figure composed of six panels arranged in two rows, showing line graphs of differential emission measure versus temperature (top row) and two-dimensional emission measure with an overlaid emission line (bottom row).}
    \label{fig:synthetic_dems}
\end{figure*}

We evaluate these expressions using functions from NumPy\footnote{\url{https://numpy.org/}} \citep{harris_array_2020} and SciPy\footnote{\url{https://scipy.org/}} \citep{virtanen_scipy_2020}.

Figure~\ref{fig:synthetic_dems} shows the DEM, $\xi(T)$, and the two-dimensional emission measure distribution, $\Xi(T,v)$, for the selected snapshot of the simulated atmosphere. The distribution of plasma in temperature and velocity space is shown, with the resulting Fe~XII~195.119~\AA\ synthesised emission line overlaid. The temperature range of the plasma which is responsible for $\sim$95\% ($\pm$2$\sigma$) of the line intensity for this transition, calculated using the contribution function generated by CHIANTI, is shown between two vertical dashed lines for each panel. The three columns show the pixels with the \(\mu - \sigma\), \(\mu\) and \(\mu + \sigma\) intensities, when considering the total wavelength-integrated intensities from the Fe~XII~195.119~\AA, the blended Fe~XII~195.179~\AA, and the most intense surrounding background lines. To avoid edge effects and to sample representative interior pixels closer to the magnetic polarities, the intensity distribution used to select these representative pixels is computed after excluding a border region equal to 20\% of the shortest map side; the three representative pixels are then chosen from the remaining area.

\begin{figure*}
    \centering
    \includegraphics[width=\linewidth]{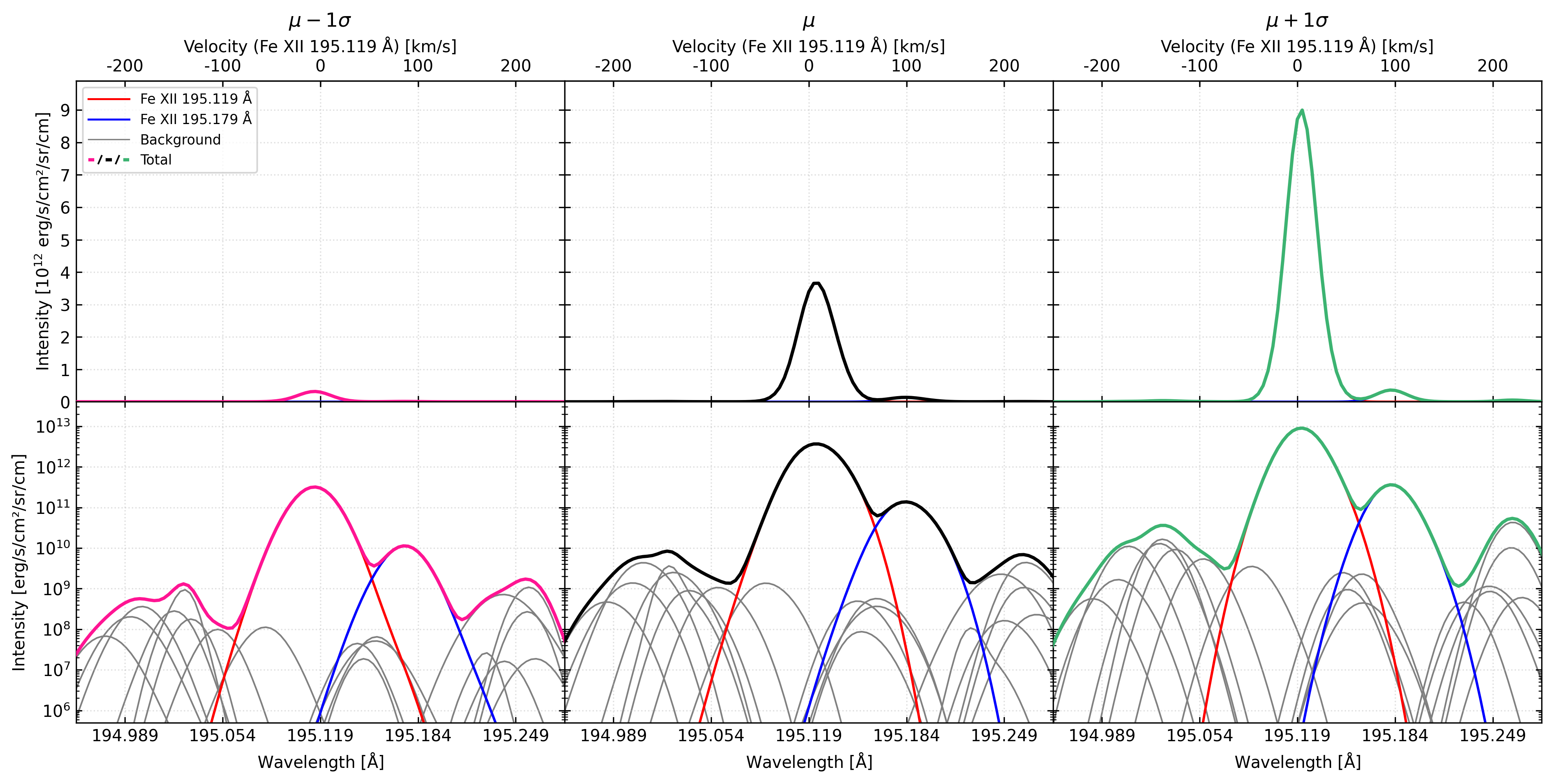}
    \caption{Synthetic emission line spectra for the three representative pixels analysed in Figure~\ref{fig:synthetic_dems}. Each panel shows the Fe~XII~195.119~\AA\ line (red), the blended Fe~XII~195.179~\AA\ line (blue), background emission from other spectral lines (grey), and the total synthetic spectrum. The top axis displays the corresponding Doppler velocity scale for the Fe~XII~195.119~\AA\ transition. The spectra are shown with a linear intensity scale in the top row and logarithmic scale in the bottom row.\\
    Alt text: Figure composed of six panels in two rows, each showing spectral line plots with wavelength on horizontal axis and intensity on vertical axis.}
    \label{fig:synthetic_spectra}
\end{figure*}

\begin{figure}
    \centering
    \includegraphics[width=\linewidth]{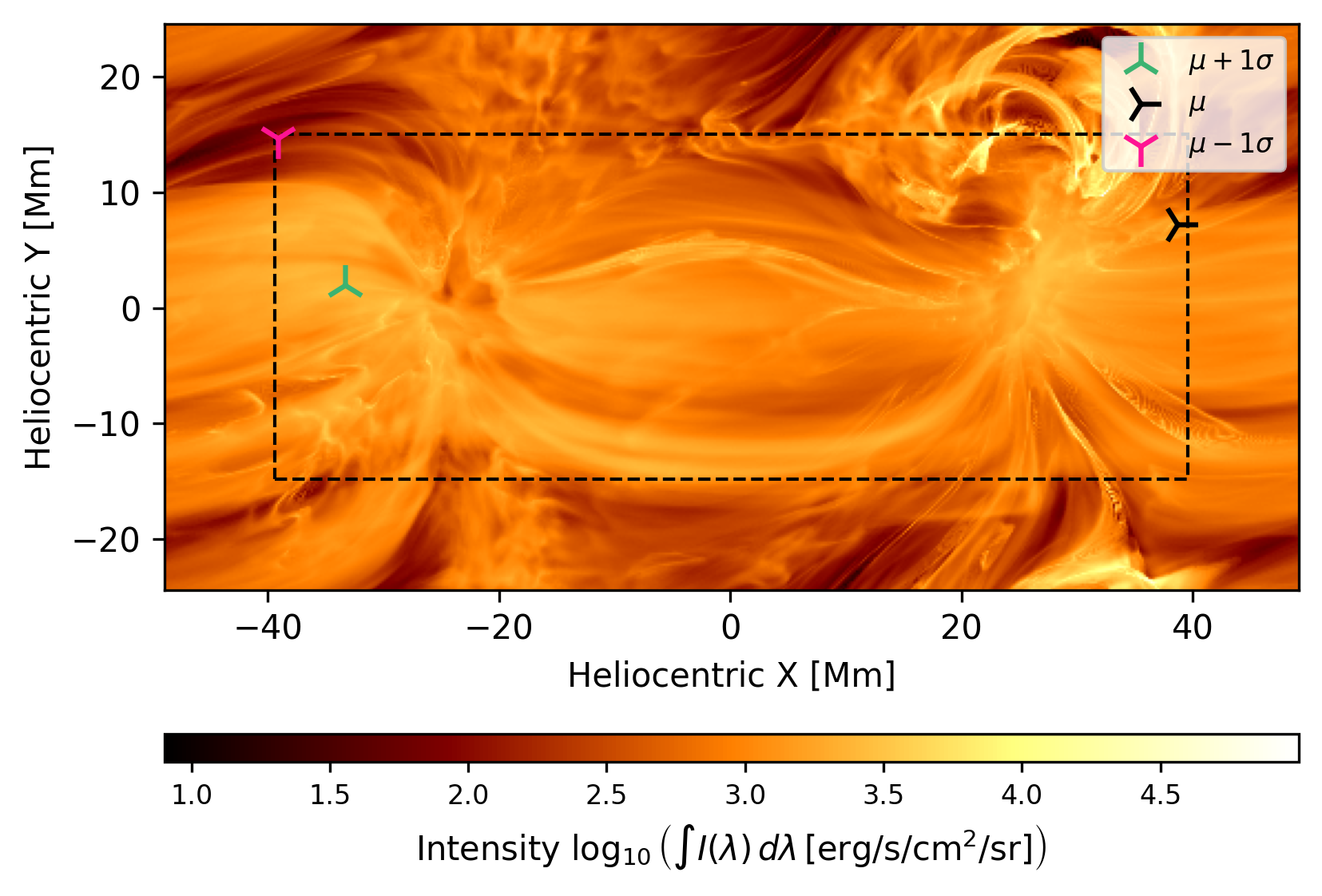}
    \caption{Total intensity of the Fe~XII~195.119~\AA, Fe~XII~195.179~\AA, and background lines for the selected snapshot. The pixels with the mean minus one standard deviation (\(\mu - \sigma\)), mean (\(\mu\)), and mean plus one standard deviation (\(\mu + \sigma\)) intensities within 20\% of the shortest side of the map to the edge are shown. This bounding region is shown with a thin dashed black line. Solid contours are drawn around the radial magnetic field, $B_z$, at the photosphere (see text for details) in white (+400~G) and black (-400~G), showing the main active region polarities (0~Mm Heliocentric~Y) and emerged destabailising parasitic polarities (+10~Mm Heliocentric~Y). Size thresholding is used to show only the large strong polarities.\\
    Alt text: Single intensity map with colour scale showing spatially-resolved emission, overlaid with white and black contour lines and three marked pixel locations.}
    \label{fig:synthetic_intensity_map}
\end{figure}

The resulting synthetic spectra for these emission lines in the three representative pixels are presented in Figure~\ref{fig:synthetic_spectra}, and the spatially-resolved total intensity of these lines is shown in Figure~\ref{fig:synthetic_intensity_map}, where the synthesised values are in agreement with those expected from observation \citep[e.g.,][]{brown_wavelengths_2008}. We also show the main polarities of the radial magnetic field at the photosphere taken at 7.5~Mm from the base of the MHD simulation, the location of the photosphere \citep{cheung_comprehensive_2018}.

The two-dimensional Differential Emission Measure (2D DEM) approach described above provides a physically motivated method for synthesising emission lines from complex, three-dimensional simulations of the solar atmosphere. By binning the plasma properties into discrete intervals of temperature and line-of-sight velocity, we construct a continuous distribution function, $\Xi(T, v)$, that captures the full range of thermal and dynamic states present along each line of sight. This approach is essential for line profiles integrated along the line-of-sight because the underlying simulation is itself discretised into finite voxels, each representing an average plasma state over a small but nonzero spatial volume. Treating each voxel as a delta function in temperature and velocity can artificially exaggerate the discreteness of the simulation, leading to unrealistic, non-physical emission line profiles composed of isolated, narrow components. Instead, by aggregating the emission measure into adjacent bins in temperature and velocity space, we recover a smooth, continuous distribution that more accurately reflects the true, unresolved sub-voxel variations in the plasma. This method assumes that the simulation's distributions of temperature, density, and velocity are statistically representative of the real solar atmosphere at the relevant spatial scales, and that the bin widths are chosen fine enough to resolve the key features of the emission line profiles, but coarse enough to avoid over-interpreting the simulation's numerical granularity. See the Appendix for more details.

In this synthesis we only consider the thermal broadening and the bulk velocity along the line of sight. Any sub-grid or turbulent motions below the simulation resolution are not captured. Such unresolved dynamics could be included by introducing an additional broadening term, assumed either constant or parameterised as a function of temperature or formation height. The divergence of the velocity field could also be used to estimate a local broadening term, but this requires assumptions about how energy cascades below the grid scale. As we are not concerned with such broadening in this study, we leave such extensions of the synthesis method for future work.

\subsection{Instrument response}\label{sec:method_response}

As we have now synthesised the emission of the atmosphere we can simulate the response of the instrument in detecting this emission.

\begin{figure*}
    \centering
    \includegraphics[width=.95\linewidth]{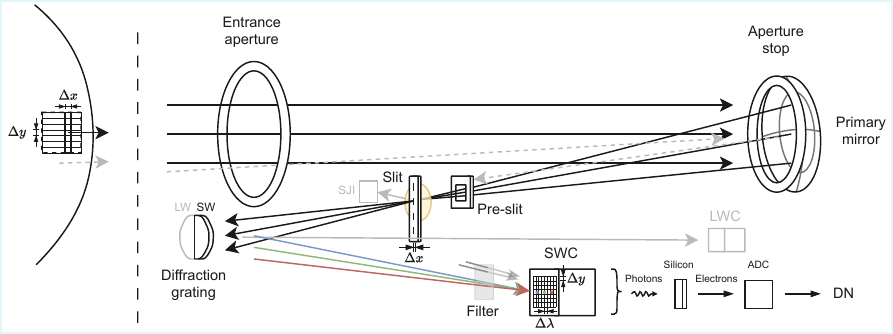}
    \caption{Schematic diagram of the SOLAR-C/EUVST short-wavelength channel (red/green/blue). Light enters through the entrance aperture and is focused by the primary mirror. Light from outside the field of view is practically fully absorbed by the pre-slit (dashed grey). The slit selects the spatial region to be observed, and the diffraction grating disperses the light spectrally. The thin-film aluminium filter blocks visible stray light (solid grey) while transmitting EUV radiation. The silicon CCD detector of the short wavelength camera (SWC) records the dispersed spectrum using an analogue-to-digital converter (ADC). We also show the slit-jaw imager (SJI) and the long-wavelength camera (LWC).\\
    Alt text: Schematic diagram showing optical path with labelled components including mirror, slit, grating, and detector.}
    \label{fig:cartoon}
\end{figure*}

\subsubsection{Instrument overview}

SOLAR-C will be launched into a Sun-synchronous orbit around Earth, meaning it will have near-continuous visibility of the solar disk. A large off-axis parabolic telescope constitutes the majority of the volume of the spacecraft, feeding light to the spectrograph and a slit-jaw imager which will provide context imaging as is done on IRIS. In this paper, we only concern ourselves with the telescope and spectrograph. Other components of the spacecraft are discussed in \cite{shimizu_solar-c_2020}, and further details of the telescope and spectrograph optical design can be found in \cite{kawate_concept_2019}, \cite{kawate_sensitivity_2020} and \cite{tsuzuki_modification_2024}.

We present a cartoon layout of the SOLAR-C/EUVST spectrograph and the SW camera in Figure~\ref{fig:cartoon}. The way the telescope and spectrograph work is as follows:

\begin{enumerate}
    \item The front aperture of the SOLAR-C telescope collects light, of all wavelengths, radiating from the Sun.
    \item As light passes through the telescope an aperture stop before the primary mirror defines a distinct aperture, through which collected photons pass and by which any others are practically all absorbed.
    \item The surviving light then reflects from the primary mirror, which directs it towards the slit plane.
    \item A pre-slit heat dump is used to absorb any photons originating from outside the field of view of the spectrograph or slit-jaw context imager, and pass in-field-of-view photons to the slit.
    \item A slit assembly, containing multiple slits with different widths, is mechanised to adjust the spatial sampling in the Solar-X direction. Some light from a spatial column on the Sun passes through the slit, corresponding with the width of the slit, with the remaining light reflected into the slit-jaw context imager.
    \item The light passed through the slit is then reflected by a diffraction grating, which disperses the light according to wavelength. This is a split diffraction grating, delivering one half of the incident beam to the SW camera and the other half to the LW camera.
    \item For the SW channel, the dispersed light passes through a thin-film aluminium optical filter which is used to block any out-of-band stray light, primarily at visible wavelengths, from passing further into the optical system. The EUV light passes through. There is some modulation of the beam in the filter plane, as an opaque mesh is needed to support the very thin optical filter. Additional optical baffling further reduces stray light.
    \item The light reaching the SW camera is absorbed by a silicon CCD detector. At this point, the incident photons liberate electrons from the silicon, which are read out and converted to data numbers (DN) for transmission to Earth.
    \item The primary mirror is intermittently driven along the Solar-X direction to scan (raster) the slit across a region on the Sun, and spatially and spectrally resolve the plasma in the solar atmosphere.
\end{enumerate}

We consider in the following text the instrument parameters dictating the performance of the spectrograph.

\subsubsection{Shot noise}

Shot noise (photon noise) is a fundamental source of statistical uncertainty arising from the discrete nature of photon generation, and is governed by Poisson statistics. The number of photons detected in a given time interval fluctuates around the mean expected value, with a variance equal to the mean. In our simulations, we account for shot noise by generating Poisson-distributed random variates for the number of detected photons in each pixel and wavelength bin, based on the synthesised photon flux.

\subsubsection{Ideal spatial and spectral sampling}

\paragraph{Slit width}

The slit width defines the spatial region on the Sun in the scanning direction (Solar-X) sampled by the spectrograph at each exposure, set by the physical width of the entrance slit in the focal plane (Figure \ref{fig:cartoon}). In our modelling, the slit transmission profile is assumed to be a top-hat function, meaning all light within the nominal slit width is transmitted equally, and light outside is blocked. In reality, the slit edges are not perfectly sharp. The finite thickness of the slit and diffraction at the edges will produce a non-ideal, slightly rounded or sloped profile. Once the slit has been manufactured the true transmission function will be measured and incorporated into our model. For SOLAR-C/EUVST, several science slit widths are available (0.2~arcsec, 0.4~arcsec, 0.8~arcsec, 1.6~arcsec), selectable by a moving slit mechanism. Light not passing through the slit is reflected to a slit-jaw imager for context imaging, as is done on IRIS.

\paragraph{Spatial sampling (plate scale)}

The spatial sampling of an instrument, also known as the plate scale, defines the relationship between a physical distance on the detector and the corresponding angular size on the sky. It is determined in the ideal case by the optical configuration and the pixel size of the detector, as seen in Figure \ref{fig:cartoon} where the focusing optics and detector pixel size dictate \(\Delta{}y\). For the SOLAR-C/EUVST short-wavelength (SW) channel the plate scale is fixed by the instrument design to 0.159~arcsec/pixel. In the case of EUVST, which is a slit spectrograph, this plate scale applies along the slit direction (Solar-Y), and is practically independent of the slit width selected for observations. \psfupdate{The spatial sampling in the along-slit direction can be adjusted by binning pixels. This technique is useful in that it improves the signal of a measurement and reduces the required exposure times.
}

\paragraph{Spectral sampling}


The spectral sampling of a spectrograph defines the wavelength interval covered by each pixel on the detector and in the ideal case is fixed by the optical design and pixel size in a similar way to the plate scale (Figure \ref{fig:cartoon}). For the SOLAR-C/EUVST short-wavelength (SW) channel, the spectral sampling is 16.9~m\AA{}/pixel, resulting in a spectral resolution of $\frac{\lambda}{\Delta\lambda}$\textgreater{}4000.

\paragraph{Implementation}

We first resample our ideal synthesised spectra to the spectral sampling of the short-wavelength detector using the integrated flux-conserving interpolation of \citet{carnall_spectres_2017} as implemented by \citet{the_astropy_collaboration_astropy_2022}, where the resampled flux $f_{\lambda_j}$ at wavelength $\lambda_j$ is given by

\begin{equation}
    f_{\lambda_j} = \frac{\sum_{i} P_{ij} w_i f_{\lambda_i}}{\sum_{i} P_{ij} w_i}\quad{}[\text{erg s}^{-1}\text{ cm}^{-2}\text{ sr}^{-1}\text{ cm}^{-1}]
\end{equation}

\noindent{}and where $f_{\lambda_i}$ is the input flux at wavelength $\lambda_i$, $w_i$ is the original bin width, and $P_{ij}$ is the fractional overlap between input and output bins.

We then use the flux-conserving spherical polygon intersection method from \texttt{Astropy} \citep{the_astropy_collaboration_astropy_2022} to resample the $x$ and $y$ spatial directions of our synthetic data to match the instrument slit width and plate scale, respectively.

\subsubsection{Point-Spread Function (PSF)}

A PSF describes how an imaging system spreads a single point of light, and is generated by many combined effects. For scanning slit spectrometers like we consider here, at the focal plane the CCD's $x$-direction corresponds to the spectral ($\lambda$) axis, while the $y$-direction corresponds to the spatial axis along the slit ($y$). See Figure \ref{fig:cartoon}. As the PSF spreads light in both directions, in the resulting data it provides blurring to spectral lines (known as instrumental broadening) and in the along-slit spatial direction (reducing the effective spatial resolution in that axis).

For SOLAR-C/EUVST, the PSF will largely be shaped by factors in the focusing optics such as the surface figure error of the primary mirror and grating and any defocusing caused by displacement of components from their ideal locations. Additionally, the solid mesh supporting the thin-film aluminium filter introduces a complex diffraction pattern of cross-shaped intensity maxima (see e.g. \cite{lin_diffraction_2001}). The current optical design of SOLAR-C/EUVST-SW predicts a PSF from the focusing optics with a full width at half maximum (FWHM) of approximately \psfupdate{0.42~arcsec (spatial) and 43~m\AA{} (spectral)}
. We include this in our results here, modelled as a Gaussian function with the appropriate width \(\sigma\), where \(\text{FWHM} = 2\sigma\sqrt{2\ln2}\). This does not consider additional PSF wings, created by a high-frequency surface figure error and the microroughness of the primary mirror and the grating, which is a very wide-angle affect. We have separately analysed the PSF expected from the thin-film aluminium filter supporting mesh using Ansys Zemax OpticStudio 2024, Release 1.03. The mesh is approximately 250~mm from the CCD, with a pitch of 390~\textmu{}m and bar width of 40~\textmu{}m, rotated 45~degrees with respect to the CCD pixel axes. We find the resulting intensity modulation is less than 1\% and so for the active region case considered here, we neglect this mesh diffraction PSF. However, in future work, particularly for limb flare studies, where the intensity contrast between the diffraction maxima and the background (with little or no plasma) is much higher and the diffraction effects may become significant, we will include this effect. The true PSF will be fully characterised during ground calibration.


\subsubsection{Optical throughput}

Optical throughput quantifies the fraction of incident photons that are captured by the telescope and successfully pass through the optical system to reach the detector, excluding detector-specific detection effects like quantum efficiency. For SOLAR-C/EUVST-SW, the optical throughput is determined by the product of several key components, expressed using $T_{\mathrm{opt}}$, as



\psfupdate{
\begin{equation}
      T_{\mathrm{opt}}\left(\lambda\right) = A_{\mathrm{ap}} \times R_{\mathrm{PM}}\left(\lambda\right)\times R_{\mathrm{DG}}\left(\lambda\right) \times T_{\mathrm{F}}\left(\lambda\right)\quad[\mathrm{cm}^2],
      \label{equ:optical_throughput}
\end{equation}
}

\noindent{}where $A_{\mathrm{ap}}$ is the aperture area (defined by the aperture stop), 
$R_{\mathrm{PM}}\left(\lambda\right)$ is the primary mirror reflectivity, 
$R_{\mathrm{DG}}\left(\lambda\right)$ is the diffraction grating reflectivity, and $T_{\mathrm{F}}\left(\lambda\right)$ is the filter transmission, all as a function of wavelength. Additional contamination factors apply, but we consider these with the filter transmission; see following text.
\psfupdate{The reflectivity of the primary mirror is a function of its multilayer coating and microroughness. The former is used to enhance reflectivity across the wavelengths of interest, and the latter is small-scale surface irregularities on optical components that scatter incident light away from the specular reflection direction. The reflectivity attenuation of the primary mirror due to microroughness is given by

\begin{equation}\label{equ:microroughness}
  R_{\mathrm{MR}} = \exp\!\left[-\left(\frac{4\pi\sigma_{\mathrm{MR}}}{\lambda}\right)^2\right]
\end{equation}

\noindent{}where $\sigma_{\mathrm{MR}}$ is the root-mean-square (RMS) surface roughness \citep[e.g.,][]{bennett_relation_1961}. We include this factor in our primary mirror reflectivity factor directly.
We see in this relationship that microroughness is particularly important in EUV telescopes as the scattering losses scale inversely with the square of the wavelength.

For the SOLAR-C primary mirror, the microroughness is nominally 0.3~nm~rms, which at 195~\AA{} corresponds to approximately a 
96\% reflectivity. 
It should be noted, however, that the 
effect of microroughness may 
deviate from the form given by Equation~\ref{equ:microroughness}
due to the multilayer coating
. This will be assessed in more detail during the manufacturing and test campaign.
}

It is important to note that optical throughput, as defined here, does not include the detector quantum efficiency (QE) as we consider detector effects later in our pipeline; the effective area (EA) is obtained by multiplying the optical throughput by the QE.

\paragraph{Thin-film aluminium filter}

The thin-film aluminium filter is designed to block visible light photons that are collected by SOLAR-C which we do not want to detect but to which the SOLAR-C/EUVST-SW CCD detector is highly sensitive, while transmitting EUV photons.

The SOLAR-C/EUVST-SW filter is manufactured by the Luxel Corporation, and is 150~nm thick aluminium, of which 1\% of the thickness is assumed to be embedded aluminium oxide (Al$_2$O$_3$), supported by a square opaque nickel mesh. An additional Al$_2$O$_3$ layer forms naturally on the filter surfaces due to exposure to air before launch. In our simulations, we can vary the thickness of this surface oxide layer to consider various cases, but for this work adopt 4~nm on each side, which is the value at which the oxide layer is expected to saturate, in the worst-case. The transmission of Al and Al$_2$O$_3$ is calculated using the work of \cite{powell_thin_1990} and \cite{henke_x-ray_1993}. The supporting nickel mesh adds an additional blocking factor of $\sim$20\%. The resulting EUV transmission of the filter ($T_F(\lambda)$) at the wavelengths we consider in this study is around 50\%.

We also consider molecular contamination on the filter surfaces. Two main classes of contaminants are the most important for space instruments, water and hydrocarbons. In our case, water molecules desorb from materials in the spacecraft in vacuum and react with aluminium, thickening the Al$_2$O$_3$ layer. As previously mentioned, we consider this saturated during ground activities and so do not concern ourselves with this desorption here. Hydrocarbons, however, deposit as carbon-rich films on optical surfaces like our filter, providing an additional throughput loss. They originate from machining residues and from the outgassing of adhesives and structural materials, and are mitigated by carefully selecting low-outgassing materials and by the thorough vacuum bakeout of components to drive off water, solvents, and other volatiles. Nevertheless, residual hydrocarbons are difficult to eliminate entirely, and we include in our model a uniform carbon contamination layer of 2~nm thickness on each side of the filter, consistent with SOLAR-C/EUVST contamination control expectations.

\subsubsection{Stray light}

Stray light refers to unwanted photons, particularly from visible wavelengths, which reach the detector and are recorded erroneously as EUV signal (see e.g. \cite{young_scattered_2022}). Full-telescope stray light analysis has been performed to quantify the amount of visible light expected at the CCD detector \citep{ishikawa_stray_2025}, the details of which will be published in full elsewhere. In such analysis, visible light unintended for the detector can arrive there through scattering from optical surfaces and from reflection and scattering from structural surfaces in the telescope. A stray light baffle helps protect the CCD detector from many of these stray light paths, and the aluminium filter provides further protection. Hinode/EIS used two optically-identical filters in series, including one at the front aperture of the telescope, to surpress stray light \citep{korendyke_optics_2006}. However, for SOLAR-C/EUVST, the requirement to observe a much broader spectral range means that no single filter at the front aperture can block visible light while transmitting the full EUVST wavelength range. As a result, our thin-film aluminium filter must be placed close to the SW CCD after the LW and SW channels have been divided.

The most recent analysis indicates that, after accounting for the visible light rejection of the filter and all other mitigations, the level of visible stray light at the SW detector is expected to be $<1$~ph/s/pixel \citep{ishikawa_stray_2025}. We include these with shot noise in our simulations presented here for correctness, but observe that such a low level makes no impact on science return.


\subsubsection{Detector Effects}

\paragraph{Quantum efficiency}

When an EUV photon is absorbed in the silicon of the CCD, its energy liberates electron-hole pairs, generating a measurable signal. The signal resulting from this process depends on several factors, including the QE of the detector at the wavelength of interest, and the mean energy required to generate each electron-hole pair in silicon. $\mathrm{QE}(\lambda)$ is defined as the probability that an incident photon of wavelength $\lambda$ will be absorbed and produce a collected charge carrier. For the CCD42-40 in the back-illuminated, uncoated configuration we use in SOLAR-C/EUVST-SW the QE at 195~\AA{} is expected to be around 76\% when the detector is at \nobreakdash-60\textdegree{}C. In our simulation, we apply the QE by drawing the number of detected photons from a binomial distribution, where each incident photon has a probability $\mathrm{QE}(\lambda)$ of being detected.

\paragraph{Dark Current}

Dark current is a source of signal and noise in CCD detectors that arises from thermally generated electrons within the silicon, even in the absence of incident light. These electrons accumulate in the same way as photoelectrons, leading to a background signal that can degrade image quality, especially during long exposures or at higher CCD temperatures. The dark current in the CCD detector depends on the detector temperature and for a Non-inverted Mode Operation (NIMO) CCD (as used on SOLAR-C/EUVST-SW) is defined piecewise as

\begin{multline}
  Q_{d}(T) =
  \begin{cases}
    Q_{d0} \times 122\,T^3\,\mathrm{e}^{-\frac{6400}{T}}, & 198\,\mathrm{K} \leq T \leq 300\,\mathrm{K} \\[2ex]
    Q_{d}(198\,\mathrm{K}), & T < 198\,\mathrm{K} \\[2ex]
    \text{undefined}, & T > 300\,\mathrm{K}
  \end{cases}\\
  \quad [\mathrm{e}^{-}\,\mathrm{pix}^{-1}\,\mathrm{s}^{-1}]
  \label{equ:dark_current}
\end{multline}

\noindent{}where \(Q_{d0} = 20{,}000\,\mathrm{e}^{-}/\mathrm{pix}/\mathrm{s}\) is dark current at 293~K, and \(T\) is the CCD temperature in Kelvin \citep{e2v_ccd42-40_2016}. We note that while \citet{e2v_ccd42-40_2016} clip the bottom end of the relationship at 230~K ($\sim$\nobreakdash-43\textdegree{}C), our component-level testing of the CCDs confirms the relationship down to at least 198~K ($\sim$\nobreakdash-75\textdegree{}C; in line with expectations from e.g. \citealt{harding_technology_2015}), and so we use this lower clipping temperature in our equation.

For our modelling we adopt a nominal CCD temperature of \nobreakdash-60\textdegree{}C (213~K), which is the expected in-flight value and corresponds to a dark current of approximately 2~e\(^{-}\)/pix/s. Like the photon signal, the dark current is also subject to Poisson noise because it arises from a discrete, random process of thermal electron generation following Poisson statistics. This parameter is variable in the code to allow testing of hot CCD scenarios against science requirements.


\paragraph{Fano Noise}

Fano noise is an intrinsic source of statistical fluctuation in the number of electrons generated when a photon is absorbed in a semiconductor detector such as silicon. Unlike pure Poisson statistics, where the variance equals the mean, the Fano effect arises because the energy required to produce each electron-hole pair is not strictly constant, but exhibits correlated fluctuations due to the physics of energy loss processes in the material.

The Fano factor, \(F\), quantifies this reduction in variance relative to the Poisson expectation. For silicon, we adopt the theoretical value \(F = 0.115\) \citep{alig_scattering_1980}. The variance in the number of electrons generated per absorbed photon of energy \(E\) is then:

\begin{equation}
  {\sigma_{\mathrm{Fano}}}^2 = F \cdot N_e = F \cdot \frac{E}{w(T)}\quad [{\mathrm{e}^{-}}^2],
\end{equation}

where \( w(T) \) is the mean energy required to produce one electron-hole pair in silicon as a function of temperature \( T \). Empirically, \( w(T) \) can be approximated by

\begin{equation}
  w(T) = 3.71 - 0.0006 \times (T - 300) \quad \left[\mathrm{eV}\right],
\end{equation}

where \( T \) is in Kelvin. For a CCD temperature of \nobreakdash-60\textdegree{}C, this gives \( w(213\,\mathrm{K}) \approx 3.76\,\mathrm{eV} \). For a 195~\AA{} photon, this corresponds to about 17 electrons generated per photon.

In our simulations, for each photon absorbed, the number of electrons generated is drawn from a normal distribution with mean \( N_e = E/w(T) \) and variance \( {\sigma_{\mathrm{Fano}}}^2 \).

\paragraph{Read-Out Noise}


Read-out noise is a source of uncertainty in CCD detectors that arises during the process of converting the accumulated charge in each pixel into an analog signal at the input to the ADC. This noise is primarily generated by the on-chip amplifier and associated electronics as the charge is transferred and measured, and is independent of the exposure time or incident photon flux. Read-out noise is characterised by a Gaussian distribution with a standard deviation specified in rms~electrons per pixel. For the SOLAR-C/EUVST SW detector, the standard deviation of the read-out noise from the CCD is expected to be at most 11~e\(^{-}\)/pix, based on the detector specifications. In our simulations, we conservatively adopt this worst-case value of 11~e\(^{-}\)/pix. This noise is added as a Gaussian random variable to each pixel during the signal processing stage. Small additional read-out noise contributions are expected from the analogue electronics. These have not yet been measured and are expected to be small in comparison, and so are neglected here.

\paragraph{Electronic Gain and Quantisation}

The electronic gain of a CCD detector is the conversion factor between the number of electrons collected in each pixel and the data number (DN) recorded. It is defined as the number of electrons per DN (\(\mathrm{e}^-/\mathrm{DN}\)), and is set by the on-chip charge-to-voltage amplifier, readout electronics analog gain and ADC volts-to-DN conversion. For the 16-bit Analogue Devices AD7961 ADC on SOLAR-C/EUVST-SW, the maximum DN it can record is 2$^{16}$ \citep[65,535~DN;][]{analog_devices_ad7961_2013}. A gain ($G$) is chosen to make sure the largest signals expected almost reach this value, and in our case this results in \(G = 2.78\,\mathrm{e}^-/\mathrm{DN}\). The ADC conversion of electrons to DN results in an integer number of DN in a process called quantisation. The AD7961 ADC we use performs mid-tread quantisation meaning that after the gain is applied, values are rounded to the nearest integer. We do not account for differential non-linearity (DNL), which describes variations in the actual width of each DN step, as for the AD7961 the DNL is negligible compared to the detector's read-out noise.

We note that we also need to be aware of the full-well capacity of each CCD pixel, this being the most electrons each pixel can contain before being read out without electrons spilling out into neighbouring pixels (blooming). In our case this limit is expected to be around 150,000~e$^-$/pixel \citep{e2v_ccd42-40_2016}, and is compatible with the gain where both the blooming limit and ADC saturation are reached at similar signal levels. As the SOLAR-C/EUVST-SW CCD full-well capacity has not yet been measured, in our simulation code we use clipping of the DN/pixel value (ADC saturation) rather than the number of electrons/pixel (blooming).

Once the true full-well capacity of the CCD pixels is measured, the gain may be adjusted.

\subsubsection{Measurement accuracy}\label{sec:measurement_accuracy}

Given the various noise sources described above, we use a Monte Carlo approach to estimate the precision of our Doppler velocity measurements. For each iteration, we simulate the full instrument response with all relevant noise sources to create a noisy line profile, then fit it. \psfupdate{We use a two-component Gaussian fit with peaks offset by a fixed 0.06~\AA{} to capture the Fe~XII~195.119~\AA{} and Fe~XII~195.179~\AA{} components seen in Figure~\ref{fig:synthetic_spectra}}. Since the noise is stochastic, each iteration produces a different profile from the same underlying plasma. By repeating this process many times, we determine the standard deviation of the fitted parameters, providing an estimate of measurement precision of SOLAR-C/EUVST-SW.

SOLAR-C/EUVST implements a mechanism on the primary mirror capable of correcting for any spacecraft thermally-induced spectral line shifts. This is expected to be practically 100\% effective. The calibration of the 0~km/s spectral line positions on the detector will be done with observations of the quiet Sun, which we can also consider to be effective (see e.g. \citealp{young_velocity_2012} for this procedure with Hinode/EIS). Therefore, the precision we calculate SOLAR-C/EUVST-SW will reach can also be taken as the accuracy of the detector.

The statistical precision of any results we gather from our Monte Carlo simulations depends directly on the number of iterations used to sample the underlying statistical distributions. As we increase the number of Monte Carlo realisations, the standard deviation of the fitted parameter distribution converges to the true measurement uncertainty following the law of large numbers. The convergence rate scales approximately as $1/\sqrt{N}$, where $N$ is the number of iterations. 
\psfupdate{For all results presented in this paper, we employ 500 Monte Carlo iterations per pixel (corresponding to a statistical precision of $\sim$4\%).}

\subsubsection{Instrument response pipeline}
To summarise, our method converts the temperature, density, and velocity from a 3D MHD-simulated solar atmosphere into optically thin solar emission, and then into a realistic instrument response. Our instrument response pipeline does this by applying the above described effects in the following manner (we denote noise using $\sim$):

\begin{enumerate}
  \item \textbf{Line synthesis:} We synthesise the spectral line intensity using temperature, density, and velocity distributions along the line of sight using Equation~\ref{equ:final_synth_intensity}, giving:
  
  \begin{equation}
    I_{ij}(\lambda) \quad [\text{erg}\,\text{s}^{-1}\,\text{cm}^{-2}\,\text{sr}^{-1}\,\text{cm}^{-1}].
  \end{equation}

  \item \textbf{Photon number:} We divide this new spectrum by the energy of one photon at frequency $\nu$, $E=h\nu$~[erg/photon] to calculate the number of photons per second, per area, per solid angle, per wavelength:
  
  \begin{equation}
    I_{\gamma}(\lambda)=\psfupdate{I_{ij}}(\lambda)\times{}\frac{1}{h\nu} \quad [\text{ph}\,\text{s}^{-1}\,\text{cm}^{-2}\,\text{sr}^{-1}\,\text{cm}^{-1}].
  \end{equation}

  \item \textbf{Exposure time:} We then multiply by the exposure time [s] to calculate the number of photons per area, per solid angle, per wavelength for a given exposure:
  
  \begin{equation}
    N(\lambda) = I_\gamma(\lambda) \times t \quad [\text{ph}\,\text{cm}^{-2}\,\text{sr}^{-1}\,\text{cm}^{-1}].
  \end{equation}

  \item \textbf{Optical throughput:} We then multiply by the optical throughput (in cm$^2$), this being the collecting area and optical surface throughputs excluding detector QE (Equation~\ref{equ:optical_throughput}), to calculate the number of photons per solid angle per wavelength for a given exposure:
  
  \begin{equation}
    C(\lambda) = N(\lambda) \times T_{\mathrm{opt}} \quad [\text{ph}\,\text{sr}^{-1}\,\text{cm}^{-1}].
  \end{equation}

  \item \textbf{Subtended solid angle:} We include the solid angle subtended by one detector pixel calculated using $\Omega_{pix}=\frac{\Delta{}x\times\Delta{}y}{r^2}$~[sr], where $\Delta{}x$ and $\Delta{}y$ are the slit width and plate scale respectively, and where $r^2$ is the distance from the spacecraft to the Sun. This gives the number of photons per row of CCD pixels (this being the whole spectra for this spatial position), per wavelength for a given exposure:
  
  \begin{equation}\label{equ:photons_subtended_angle}
    S(\lambda) = C(\lambda) \times \Omega_{pix} \quad [\text{ph}\,\text{cm}^{-1}].
  \end{equation}

  \item \textbf{Wavelength:} We then multiply by the spectral sampling of the instrument (in cm/pix) to determine how much of the wavelength direction of this spectra is covered by each pixel, giving the number of photons per detector pixel for a given exposure:
  
  \begin{equation}
    P(\lambda) = S(\lambda) \times \Delta\lambda \quad [\text{ph}\,\text{pix}^{-1}].
  \end{equation}

  \item \textbf{Point spread function:} We apply the PSF by performing a 2D convolution of the along-slit and spectral directions with our Gaussian PSF for each slit position. This leaves the units unchanged:
  
  \begin{equation}
    B(\lambda) = P(\lambda) \ast \text{PSF} \quad [\text{ph}\,\text{pix}^{-1}].
  \end{equation}

  \item \textbf{Quantum efficiency:} We \psfupdate{apply shot noise, then} consider the proportion of photons which will be detected by the silicon \psfupdate{through the detector QE using a binomial distribution ($\mathcal{B}$)}:
  

  \psfupdate{
  \begin{equation}
    D(\lambda) \sim \mathcal{B}\!\left(\tilde{B}(\lambda), \mathrm{QE}(\lambda)\right)
    \quad [\text{ph}\,\text{pix}^{-1}],
    \label{equ:pipeline_photons_to_qe}
  \end{equation}
  }

  \item \textbf{Fano effect:} We use the Fano effect to calculate the number of electrons which will be generated for each detected photon and include associated Fano noise, meaning we now have the number of electrons per detector pixel for a given exposure:
  
  \begin{equation}
    E(\lambda) = D(\lambda) \times \tilde{F}_{\text{Fano}} \quad [\text{e}^-\,\text{pix}^{-1}].
  \end{equation}

  \item \textbf{Stray light:} By considering the flux of visible light photons at the detector plane, and applying the QE at visible wavelengths, we add additional electrons resulting from these incident visible light photons:
  
  \begin{equation}
    E_s(\lambda) = E(\lambda) + \tilde{E}_{\text{stray}} \quad [\text{e}^-\,\text{pix}^{-1}].
  \end{equation}

  \item \textbf{Dark current:} We add a dark current signal with associated noise to this, in the form of additional electrons per detector pixel:
  
  \begin{equation}
    E_d(\lambda) = E_s(\lambda) + \tilde{E}_{\text{dark}} \quad [\text{e}^-\,\text{pix}^{-1}].
  \end{equation}

  

  \item \textbf{Detector gain:} We convert the signal of electrons per detector pixel into DN per detector pixel using the detector gain in DN/e$^-$, by rounding to the nearest integer DN value, and by clipping to the ADC saturation limit:

  \begin{equation}
    \mathrm{DN} = \min\!\left(\big\lfloor E_d(\lambda) \times G + 0.5 \big\rfloor,\, 65535\right) \quad [\mathrm{DN}\,\mathrm{pix}^{-1}]
  \end{equation}

\end{enumerate}

We assume the simulated plasma is centred at disk centre as viewed from Earth, and that curvature of the solar surface is negligible given the scale of the active region, and so integrate plasma along the $z$ direction of the simulation as the line-of-sight.

\begin{figure}
    \centering
    \includegraphics[width=\linewidth]{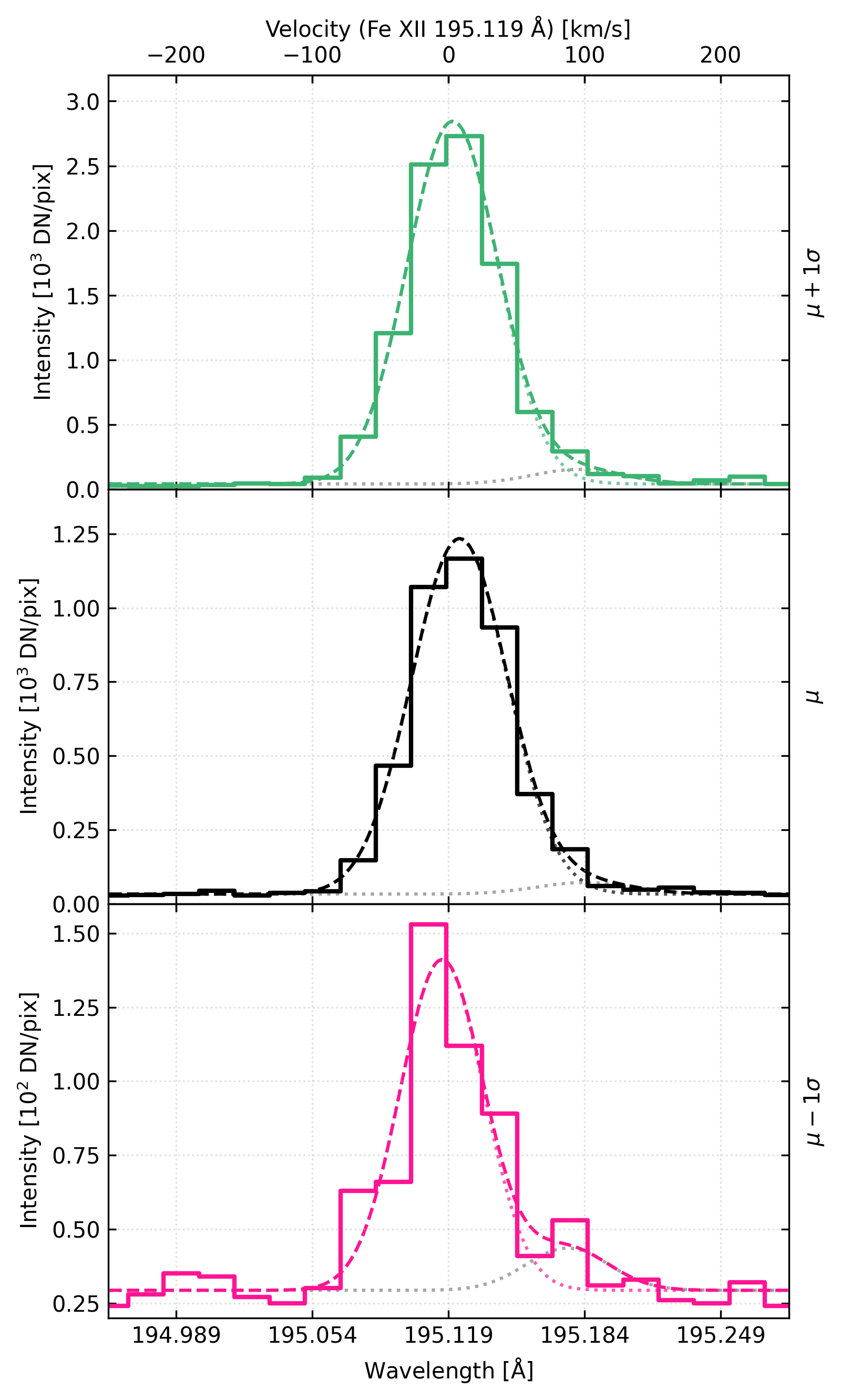}
    \caption{The simulated \psfupdate{SOLAR-C/EUVST spectral line profiles around 195.119~\AA\ }(solid lines), in DN/pix using a 40~s exposure time and the \psfupdate{0.4~arcsec} slit, for the key pixel locations in the synthesised atmosphere identified in Figure~\ref{fig:synthetic_intensity_map}. Fitted Gaussians are shown with \psfupdate{coloured} dashed lines, \psfupdate{and the two components capturing the Fe~XII~195.119~\AA\ and Fe~XII~195.179~\AA\ lines are shown with coloured and grey dotted lines respectively}.\\
    Alt text: Figure composed of three spectral line plots arranged vertically, each showing solid and dashed curves representing observed and fitted profiles.}
    \label{fig:instrument_key_pixel_spectra}
\end{figure}

An example of the final spectrum is shown in Figure~\ref{fig:instrument_key_pixel_spectra}, where we show the final number of DN we expect per detector pixel. Here, the spectrum is wavelength-calibrated so that photons from a net-zero Doppler velocity are centred on the detector pixel corresponding to the rest wavelength of the emission line. We see the line of best fit 
to the line profile well matches the centroid locations seen in Figure \ref{fig:synthetic_dems}. We note the pedestal the emission lines sit on of approximately \psfupdate{25~DN/pix}, due to the build up of dark current in the pixels over the exposure. This will likely be subtracted during level-0 to level-1 data processing, as is done with the Hinode/EIS pipeline.

\subsubsection{Comparison with Hinode/EIS}\label{sec:compare_wth_eis}

\begin{table}
  \centering
  \begin{minipage}{\linewidth}
    \centering
    \caption{Instrument parameters for Hinode/EIS and SOLAR-C/EUVST-SW where those of EIS are different from EUVST-SW and relevant to the instrument pipeline. \psfupdate{Those given for EIS are after instrument construction, while those for EUVST-SW are design and specification values.}}
    \label{tab:instrument_comparison}
    \begin{tabular}{lcc}
      \hline
      Parameter & EUVST-SW & EIS \\
      \hline
      \textbf{Optical Properties} & & \\
      Throughput\textsuperscript{*} (19.5~nm) [cm$^2$] & 1.65 & 0.47 \\
      $\Delta{}y$ [arcsec pix$^{-1}$] & 0.159 & 1.0 \\
      $\Delta{}\lambda$ [m\AA{} pix$^{-1}$] & 16.9 & 22.3 \\
      \psfupdate{Spatial PSF (FWHM) [arcsec]} & \psfupdate{0.42} & \psfupdate{3.0} \\
      \psfupdate{Spectral PSF (FWHM) [m\AA{}]} & \psfupdate{43} & \psfupdate{67} \\
      \textbf{Detector Properties} & & \\
      QE\textsuperscript{*} (19.5~nm) & 0.76 & 0.64 \\
      Read-Out Noise (RMS) [e$^-$ pix$^{-1}$] & 11.0 & 6.0 \\
      $Q_d(293 K)$ [e$^-$ pix$^{-1}$ s$^{-1}$] & 20,000.0 & 250.0 \\
      $G$ [e$^-$ DN$^{-1}$] & 2.78 & 6.6 \\
      \hline
    \end{tabular}
    \vspace{3pt}

    \begin{minipage}{\linewidth}
      \raggedright
      \textsuperscript{*}\,Different design philosophies mean the throughput and QE of EIS were optimised for a strong peak around 19.5~nm, while EUVST-SW instead focuses on high sensitivity across the whole 17-21~nm waveband. See Section~\ref{sec:compare_wth_eis} for details.
    \end{minipage}
  \end{minipage}
\end{table}

We find a comparison with Hinode/EIS useful to showcase the improvements we can expect from SOLAR-C/EUVST, and can also apply this instrument pipeline to the instrument parameters of Hinode/EIS. We take these as they were at launch, shown in Table~\ref{tab:instrument_comparison} where those parameters are different from SOLAR-C/EUVST-SW and relevant to our instrument pipeline.

The design philosophies of Hinode/EIS and SOLAR-C/EUVST-SW are different in their approach to sensitivity optimisation. Hinode/EIS uses multilayer coatings on the primary mirror and SW diffraction grating, and thinned back-illuminated CCDs, to highly optimise the throughput at the centre of the SW waveband (around 19.5~nm; the Hinode/EIS SW waveband is the same 170--210~\AA{} covered by SOLAR-C/EUVST-SW). The primary mirror and diffraction grating performance across the waveband for Hinode/EIS are shown in Figure 28 of \cite{culhane_euv_2007}, and the QE of the CCDs across the SW waveband are shown for the Hinode X-Ray Telescope (Hinode/XRT) CCDs in Figure~16 of \cite{golub_x-ray_2007}, these being identical CCDs to those used by Hinode/EIS. In the case of SOLAR-C/EUVST-SW, the instrument employs optical coatings designed to provide high sensitivity across the entire SW waveband. We therefore note that while the Fe~XII~195.119~\AA{} line shown here is good for a comparison of spatial resolutions, it does not best demonstrate the effective area improvement of SOLAR-C/EUVST-SW as it is the wavelength where the Hinode/EIS sensitivity was specifically maximized. See also \cite{young_euv_2007} for the Hinode/EIS sensitivity across its wavebands.

As we use optical throughput in our pipeline to exclude detector QE and separate the number of photons received at the detector plane from the number of photons which are detected by the CCD, we use Equation~\ref{equ:pipeline_photons_to_qe} to calculate the optical throughput for Hinode/EIS from the effective area (including QE) of 0.30~cm$^2$ given in \texttt{EIS\_EffArea\_B.003}, available in \textit{SolarSoft} \citep{freeland_data_1998}. We use a QE of 0.64, which was used to calculate this effective area and described in \cite{mariska_eis_2010}. We use a plate scale of 1.0~arcsec~pix$^{-1}$ \citep{culhane_euv_2007} and a spectral sampling of 22.3~m\AA{}~pix$^{-1}$ \citep{korendyke_optics_2006}. 
We use a symmetrical Gaussian to model the PSF \psfupdate{with a FWHM of 3~pixels \citep{ugarte-urra_eis_2010}, corresponding to 3~arcsec (spatial) and 67~m\AA{} (spectral).} We note that Hinode/EIS may have an asymmetric PSF, described in \cite{ugarte-urra_eis_2010} and with several possible supporting observations \citep[e.g.,][]{young_velocity_2012, warren_spectroscopic_2018, brooks_drivers_2020}, but find it unnecessary to consider any asymmetric shape here.

We use a readout noise of 6.0~e$^-$~pix$^{-1}$ RMS and a gain of 6.6~e$^-$~DN$^{-1}$ \citep{culhane_euv_2007}. The Hinode/EIS 42-20 CCDs use Advanced Inverted Mode Operation (AIMO), as opposed to the NIMO CCDs on SOLAR-C/EUVST, and so have a lower dark current and different relationship with CCD temperature. We use a dark current of $Q_{d0,\mathrm{EIS}}=$250.0~e$^-$~pix$^{-1}$~s$^{-1}$ at 293~K and the relationship 



\begin{multline}
  Q_{d,\mathrm{EIS}}(T) =
  \begin{cases}
    \begin{aligned}[t]
      Q_{d0,\mathrm{EIS}} \times 1.14\times10^6\,T^3\\
      \qquad{}\times \mathrm{e}^{-\frac{9080}{T}}
    \end{aligned}
    & 230\,\mathrm{K} \leq T \leq 300\,\mathrm{K} \\[2ex]
    Q_{d0,\mathrm{EIS}}(230\,\mathrm{K}),
    & T < 230\,\mathrm{K} \\[2ex]
    \text{undefined},
    & T > 300\,\mathrm{K}
  \end{cases}\\
  [\mathrm{e}^{-}\,\mathrm{pix}^{-1}\,\mathrm{s}^{-1}]
\end{multline}

\noindent{}in place of Equation~\ref{equ:dark_current}, for the Hinode/EIS AIMO 42-20 CCDs \citep{e2v_ccd42-10_2016}. We run our simulations with the Hinode/EIS CCDs at \nobreakdash-40\textdegree{}C ($\sim$233~K) \citep{bradley_warm_2025}.

\section{Results}

\subsection{Spatial and spectral resolution}\label{sec:results_spatial}

\begin{figure*}
    \centering
    \includegraphics[width=\linewidth]{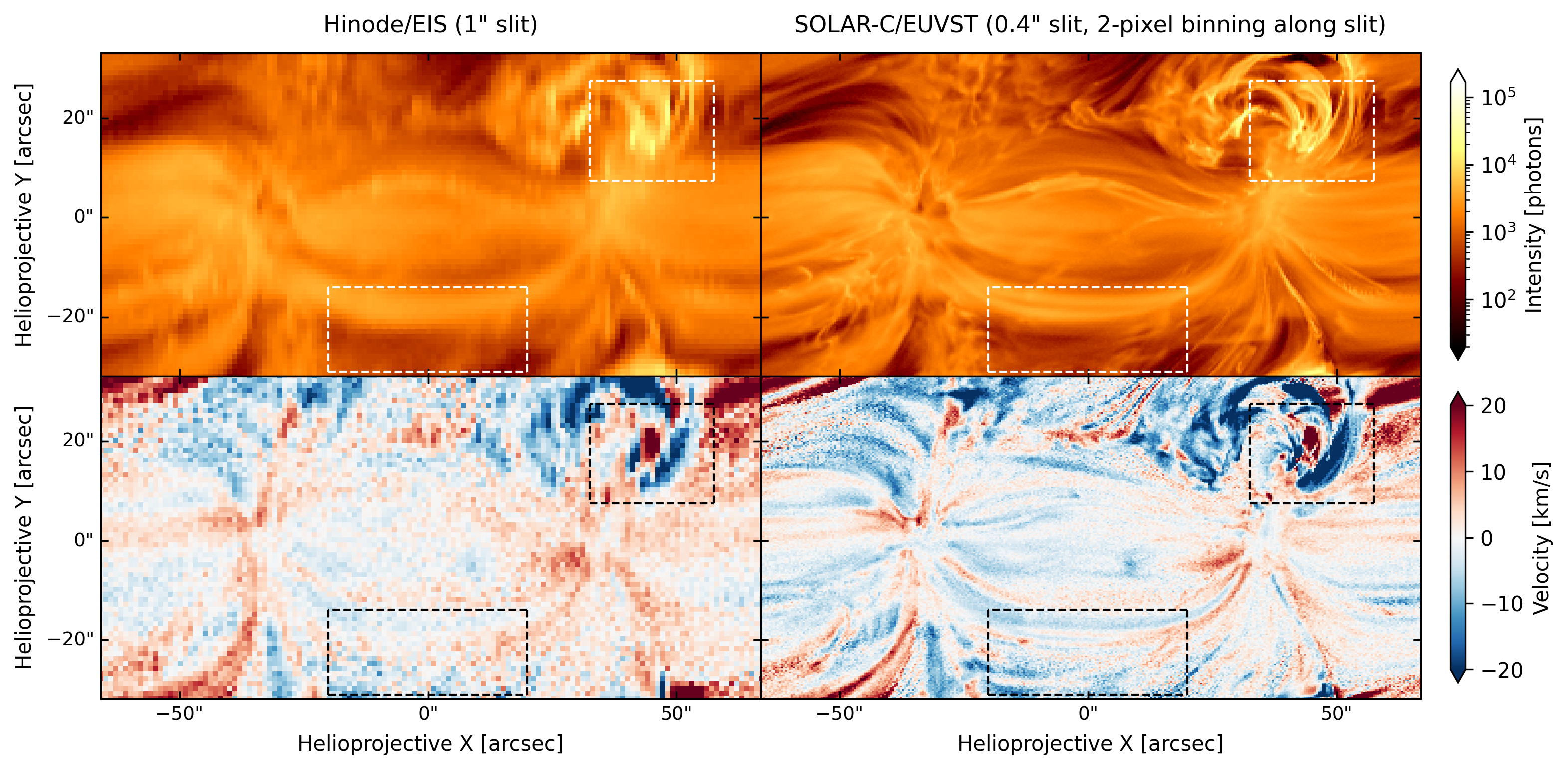}
    \caption{Comparison of simulated observations by Hinode/EIS 1~arcsec slit (left) and SOLAR-C/EUVST \psfupdate{0.4~arcsec} slit \psfupdate{with 2-pixel along-slit binning} (right) of the Fe~XII~195.119~\AA\ emission line for a 40~s exposure per slit (Helioprojective-X) position. Top row: intensity maps in units of photons at the detector plane, including point spread function and shot noise. Bottom row: line-of-sight Doppler velocity maps derived from single-component Gaussian fits to the final simulated instrument spectral profiles in DN/pix with all additional instrument noise effects included. Dashed boxes are used to show areas of closer analysis shown in Figures~\ref{fig:instrument_velocity_flare} and \ref{fig:instrument_loop}. This simulated observation uses one simulation snapshot $\sim$30~minutes before a flare, where in reality the plasma would evolve beneath the spacecraft as the active region was scanned. This is discussed further in the text.\\
    Alt text: Four-panel comparison arranged in two-by-two grid, with intensity maps in top row and velocity maps in bottom row for two instruments.}
    \label{fig:instrument_maps}
\end{figure*}

Figure~\ref{fig:instrument_maps} shows the Fe~XII~195.119~\AA{} intensity and Doppler velocity measurements of the synthetic atmosphere made by both SOLAR-C/EUVST and Hinode/EIS. We present intensity in photons/pix at the detector plane rather than DN/pix because the different dark current accumulation rates between the two instruments make direct comparison of DN values inappropriate (Table~\ref{tab:instrument_comparison}). The Doppler velocity measurements are taken by calculating the shift of the Gaussian centroid fitted to the final simulated spectral profiles in DN/pix (as in Figure~\ref{fig:instrument_key_pixel_spectra}), and so includes all noise sources described in Section~\ref{sec:method}.

In the intensity panels, we see that SOLAR-C/EUVST fully resolves the structures present in the full simulation-resolution synthesised atmosphere of Figure~\ref{fig:synthetic_intensity_map}, where Hinode/EIS does not. Notably, the multi-threaded loops connecting the two main magnetic polarities are only partially resolved in Hinode/EIS. This is sufficient for reasonable Doppler velocity measurements at larger scales and for observing the downflowing of plasma along these loops towards the active region footpoints. However, the detail of smaller-scale strands within the simulated loops is not resolved. SOLAR-C/EUVST can be seen to distinguish between these more elemental plasma flows within the loops. In the case of some active region loops, Doppler velocity reversals are seen in SOLAR-C/EUVST, indicative of siphon flows. Such detail is hard to distinguish with Hinode/EIS.



\begin{figure}
  \begin{subfigure}{\columnwidth}
    \centering
    \includegraphics[width=.87\textwidth]{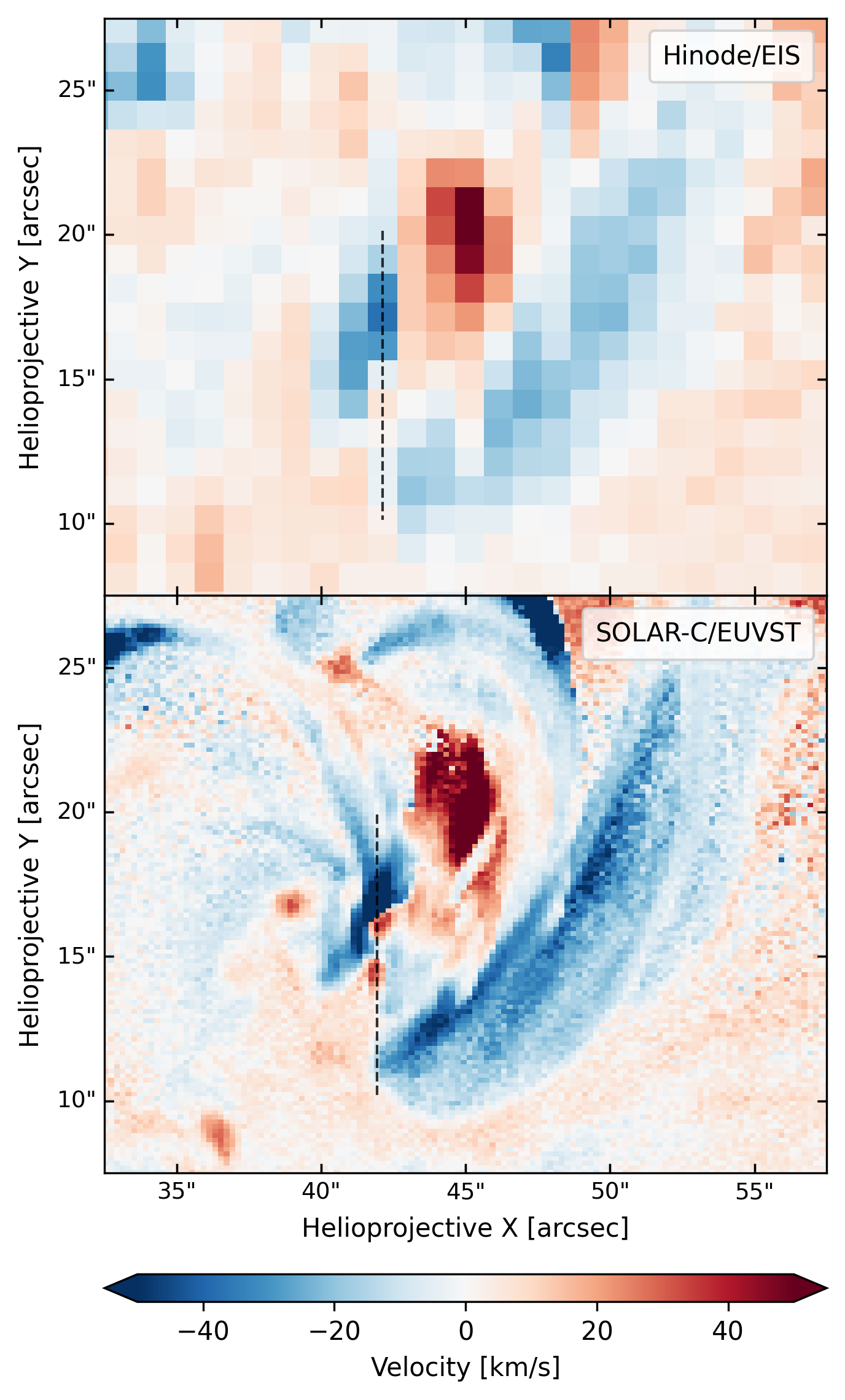}
    \caption{Pre-flare velocity maps (Hinode/EIS and SOLAR-C/EUVST).}
    \label{fig:instrument_velocity_maps_flare}
  \end{subfigure}
  
  \begin{subfigure}{\columnwidth}
    \centering
    \includegraphics[width=\textwidth]{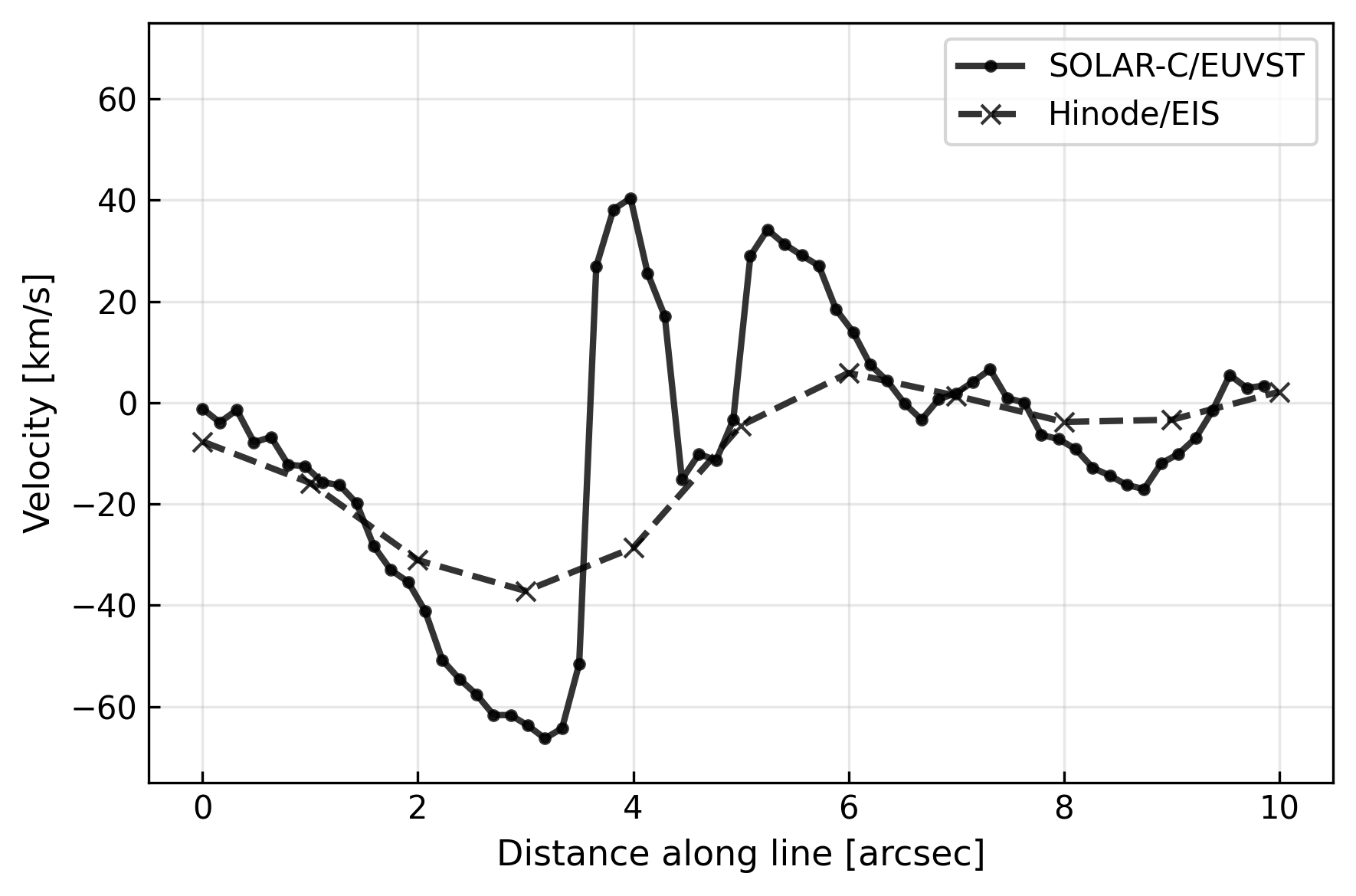}
    \caption{Velocity along the dashed line in panel~(\subref{fig:instrument_velocity_maps_flare}).}
    \label{fig:instrument_velocity_line_profile_flare}
  \end{subfigure}

  
  \caption{Comparison of simulated Fe~XII~195.119~\AA\ (log~$T\sim6.2$) Doppler velocity measurements of pre-flare coronal plasma by Hinode/EIS (1~arcsec slit) and SOLAR-C/EUVST (0.2~arcsec slit). Panel~(\subref{fig:instrument_velocity_maps_flare}) shows Doppler velocity maps for Hinode/EIS (top) and SOLAR-C/EUVST (bottom). A dashed black line is drawn over part of the atmosphere, corresponding to the Doppler velocity profile shown in panel~(\subref{fig:instrument_velocity_line_profile_flare}). Panel~(\subref{fig:instrument_velocity_line_profile_flare}) shows Doppler velocity measurements along this dashed line for the two instruments. The final simulated instrument spectral profiles for a 40~s exposure per slit position in DN/pix were fitted.\\
  Alt text: Figure (a) shows two velocity maps for the two instruments, and (b) shows a line graph comparing velocity measurements.}
  \label{fig:instrument_velocity_flare}
\end{figure}

The most striking differences between the two instruments are seen in the pre-flare region, shown in Figure~\ref{fig:instrument_velocity_flare}, where complex structure and sharp velocity gradients between strong upflowing and downflowing plasma are resolved in SOLAR-C/EUVST to approximately one detector pixel, equivalent to around 150~km at the Sun. This is best demonstrated by considering the Doppler velocity along the dashed line in Figure~\ref{fig:instrument_velocity_maps_flare}, shown in Figure~\ref{fig:instrument_velocity_line_profile_flare}. Here, we see that what is well resolved by SOLAR-C/EUVST as a sharp, approximately 300~km separation between upflowing and downflowing plasma is smoothed in Hinode/EIS, with the double peak in downflowing plasma along the line lost and reduced from the resolved $\sim$40~km/s to $\sim$5~km/s.

We also see that, at this snap shot about 30-minutes before the flare, SOLAR-C/EUVST measures the plasma flows at their simulation values of between around \nobreakdash-60~km/s and 40~km/s, whereas Hinode/EIS only measures them as between around \nobreakdash-35~km/s and 5~km/s. The range of values measured by Hinode/EIS is therefore approximately half the true range. This region of the atmosphere sits above newly emerged flux which triggers a solar flare. We discuss the potential pre-flare mechanisms SOLAR-C/EUVST is seeing in Section~\ref{sec:discussion}.

We note that taking the Doppler velocity measurements shown in Figure~\ref{fig:instrument_velocity_maps_flare} would require $\sim$125 slit positions when using the narrowest 0.2~arcsec slit. Taking the 40~seconds per slit position exposure times we use here, such an observation would take just over an hour in total, meaning that the flare would peak during the observation and also meaning that we would not \lq{}freeze\rq{} the pre-flare atmosphere at such a resolution in reality. We present an analysis of reduced exposure times and faster scanning below, including the plasma evolution we can expect to capture with different slit widths for this emission line. Furthermore, here we demonstrate only one possible observing mode of the spectrograph that highlights the highest spatial resolution slit and a wide field-of-view. Other observing modes for faster scanning, such as sit-and-stare observations, using a wider slit or sparse raster, or scanning a narrower area, are also available.

\begin{figure}
  \begin{subfigure}{\columnwidth}
    \centering
    \includegraphics[width=.95\textwidth]{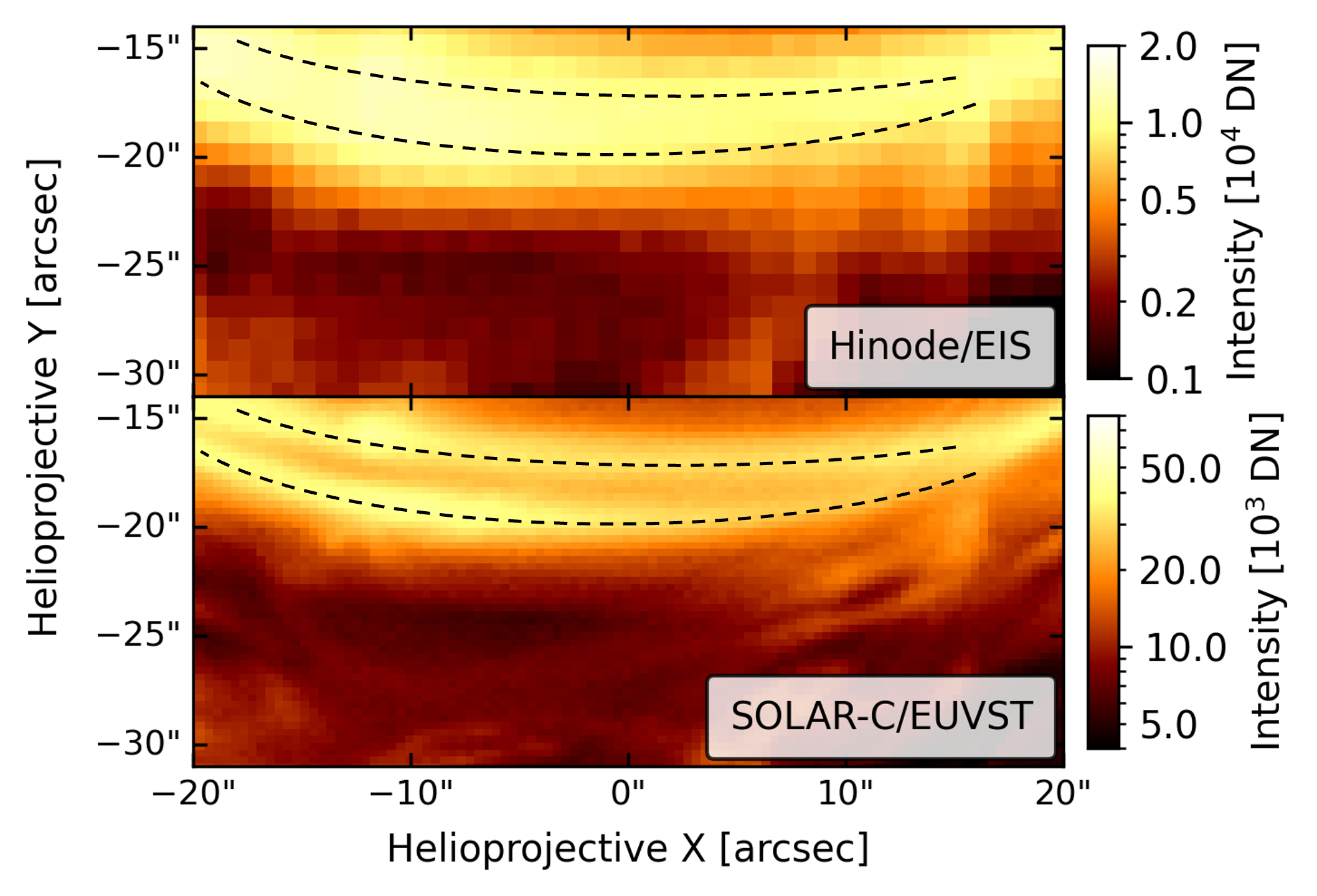}
    \caption{Loop intensity maps (Hinode/EIS and SOLAR-C/EUVST).}
    \label{fig:instrument_intensity_maps_loop}
  \end{subfigure}
  
  \begin{subfigure}{\columnwidth}
    \centering
    \includegraphics[width=\textwidth]{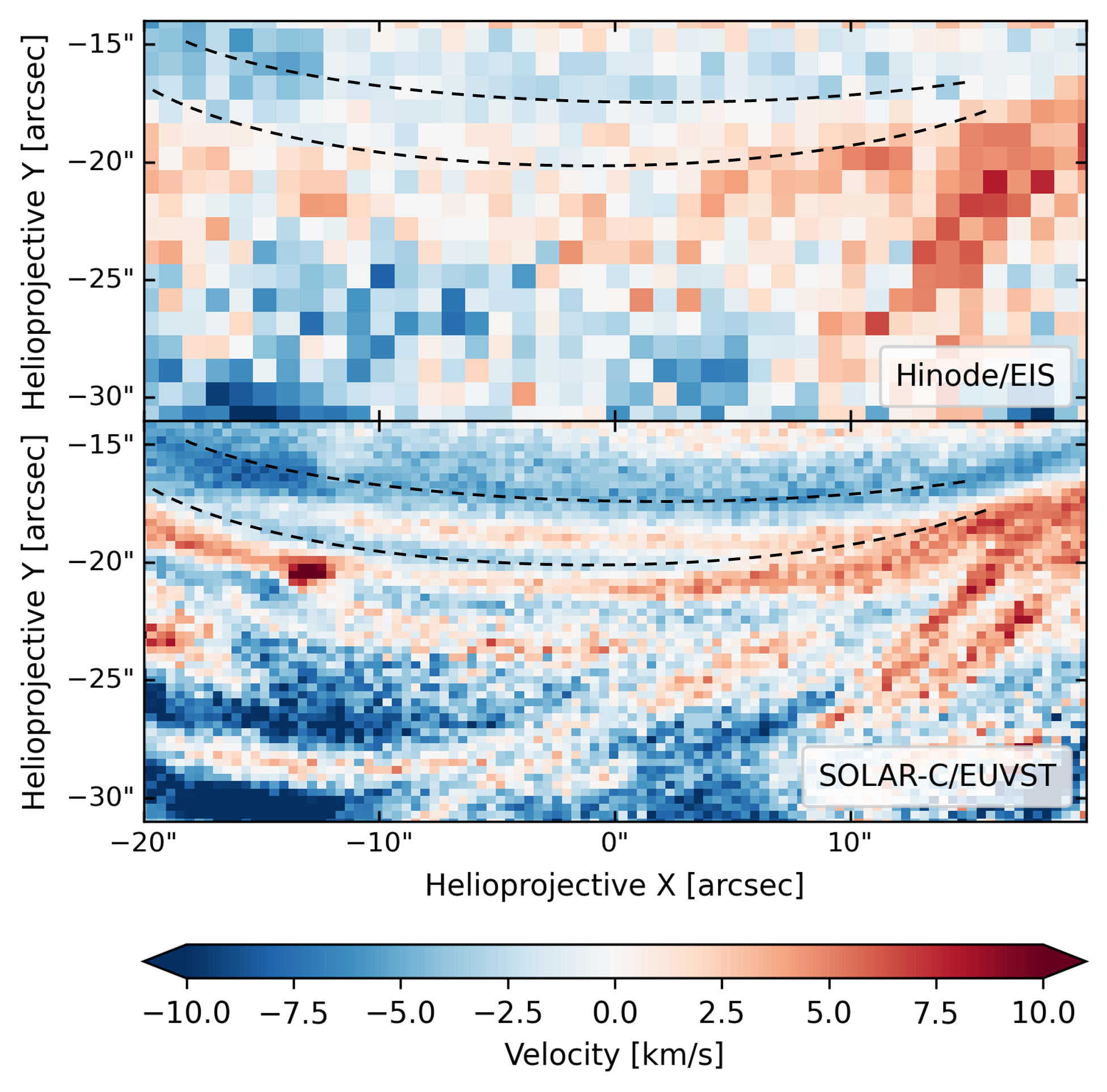}
    \caption{Loop velocity maps (Hinode/EIS and SOLAR-C/EUVST).}
    \label{fig:instrument_velocity_maps_loop}
  \end{subfigure}

  
  \caption{Comparison of simulated Fe~XII~195.119~\AA\ (log~$T\sim6.2$) intensity and Doppler velocity measurements of coronal active region loops by Hinode/EIS (1~arcsec slit; top) and SOLAR-C/EUVST (\psfupdate{0.4~arcsec slit with 2-pixel along-slit binning}; bottom). Panel~(\subref{fig:instrument_intensity_maps_loop}) shows intensity measurements. Panel~(\subref{fig:instrument_velocity_maps_loop}) shows Doppler velocity measurements. The loop strands resolved by SOLAR-C/EUVST are annotated (dashed lines). The final simulated instrument spectral profiles for a 40~s exposure per slit position in DN/pix were fitted.\\
  Alt text: Figure (a) shows two intensity maps for the two instruments, and (b) shows two velocity maps for the two instruments.}
  \label{fig:instrument_loop}
\end{figure}

We additionally consider the simulated intensity measurements made by Hinode/EIS and SOLAR-C/EUVST of the coronal loops of this active region in Figure~\ref{fig:instrument_intensity_maps_loop}. The primary southern loop as seen by Hinode/EIS is resolved as two separate distinct strands by SOLAR-C/EUVST. Resolving such structure requires not only sufficient spatial resolution but also a throughput high enough to detect the features above any noise. We see that SOLAR-C/EUVST possesses such performance in the SW detector.

There is also some structure to the lower intensity more southern region below this loop topology which Hinode/EIS partially captures, but not as completely as SOLAR-C/EUVST-SW. A question would be then, spectroscopically, whether we can still measure any flows in this region given the low signal but the appearance of some structure.



Figure~\ref{fig:instrument_velocity_maps_loop} shows the same spatial domain as Figure~\ref{fig:instrument_intensity_maps_loop}, but for the measurement of the Doppler velocity from Hinode/EIS and SOLAR-C/EUVST-SW. The latter spatially resolves some flows south of the loops in the lower intensity region. While approaching the noise level, there is clearly plasma flowing in this darker region which SOLAR-C/EUVST-SW is able to detect and Hinode/EIS is not.


\subsection{Doppler velocity measurement accuracy}

To characterise the measurement accuracy of Doppler velocities by SOLAR-C/EUVST-SW across the different intensities seen in this active region, our Monte Carlo approach allows us to consider the noise sources and instrument effects from Section~\ref{sec:method}.

\begin{figure}
  \centering
  \includegraphics[width=\linewidth]{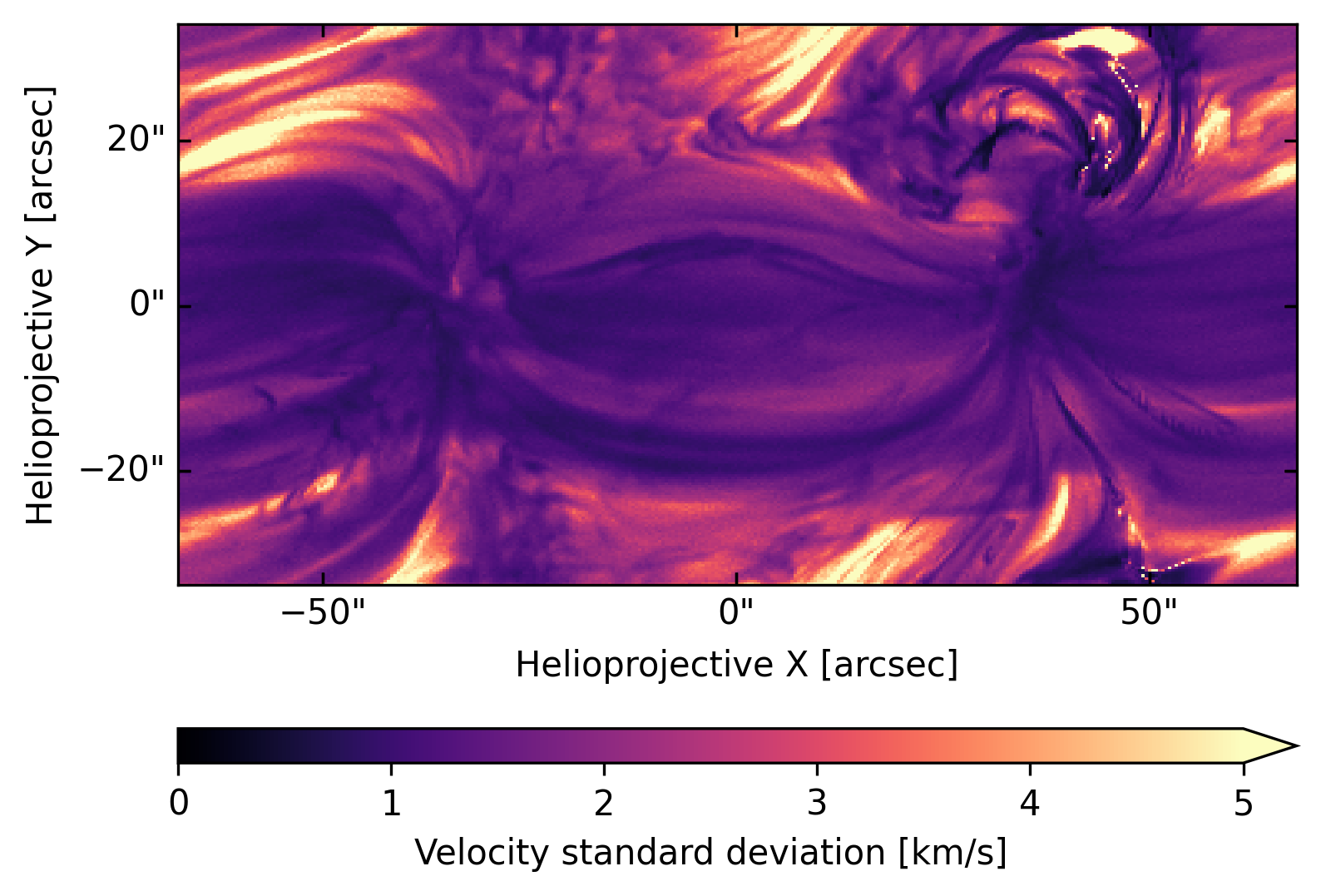}
  \caption{The standard deviation of Fe~XII~195.119~\AA\ Doppler velocity measurements made by SOLAR-C/EUVST using a \psfupdate{0.4~arcsec slit with 2-pixel along-slit binning} and 40~s exposure time, where the Doppler velocity is calculated from the centroid of a Gaussian, fitted to the simulated observation in DN/pix. 500 Monte Carlo iterations ($\sim$4\% statistical precision) were used.\\
  Alt text: Single spatial map with colour scale showing variation in measurement precision across the active region.}
  \label{fig:velocity_variation}
\end{figure}

Figure~\ref{fig:velocity_variation} shows, for the Fe~XII~195.119~\AA\ emission line for a 40-second exposure per slit position of the \psfupdate{0.4~arcsec slit with 2-pixel along-slit binning (as in Figure~\ref{fig:instrument_maps})}, the standard deviation of Doppler velocity across 500 Monte Carlo iterations ($\sim$4\% statistical precision), which we take as the precision (and accuracy; see Section~\ref{sec:measurement_accuracy}) of the Doppler velocity measurement. We see that the accuracy of the instrument for such an exposure time is well below 5~km/s for the majority of the active region with the footpoints, loops, and pre-flare region all measured to an accuracy of less than 1~km/s. There is a lower Doppler velocity measurement accuracy seen in lower intensity regions, outside of the active region core for this exposure time, approaching and exceeding 5~km/s. Such accuracies could be improved if spatial resolution \psfupdate{is sacrificed} by making observations with a wider slit \psfupdate{or increasing binning, which would improve} the signal to noise ratio of the measurements.

\begin{figure}
  \centering
  \includegraphics[width=\linewidth]{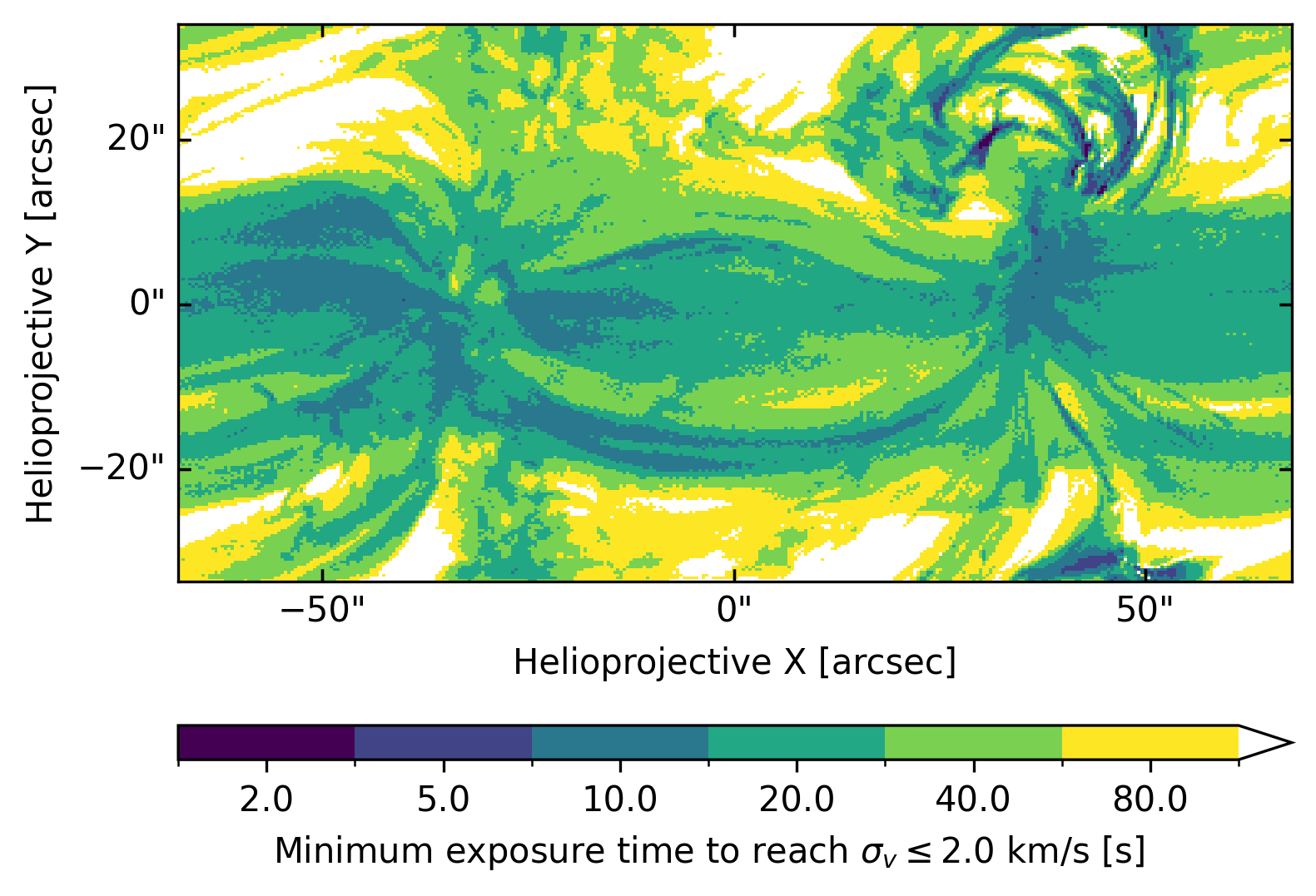}
  \caption{The minimum exposure time required to reach an Fe~XII~195.119~\AA\ Doppler velocity precision (and accuracy; see Section~\ref{sec:measurement_accuracy}) of 2~km/s, calculated from the 1$\sigma$ centroid variation in DN/pix fitted Gaussians of 500 Monte Carlo iterations ($\sim$4\% statistical precision), for SOLAR-C/EUVST using a \psfupdate{0.4~arcsec slit with 2-pixel along-slit binning}.\\
  Alt text: Single spatial map with colour scale showing required exposure times across the region.}
  \label{fig:exposure_time_requirement}
\end{figure}

In Figure~\ref{fig:exposure_time_requirement} we consider the exposure times we expect to be required to reach 2~km/s Doppler velocity precision (and accuracy) for the Fe~XII~195.119~\AA\ emission line, for the \psfupdate{0.4~arcsec slit with 2-pixel along-slit binning (as in Figure~\ref{fig:instrument_maps})}. We see that we reach this very high precision (and accuracy) at the loop footpoints within around \psfupdate{10}-seconds, and at the loops themselves in around \psfupdate{10--20}-seconds. The pre-flaring region shares similarly low exposure time requirements, with the brightest emission allowing \psfupdate{a high cadence of 2 seconds} with a measurement accuracy of 2~km/s or better. Even in moderately bright areas of the active region, exposure times of 10--20 seconds are sufficient, while only the fainter surrounding regions require exposures of 40-seconds and above.

\begin{figure*}
    \centering
    \includegraphics[width=\linewidth]{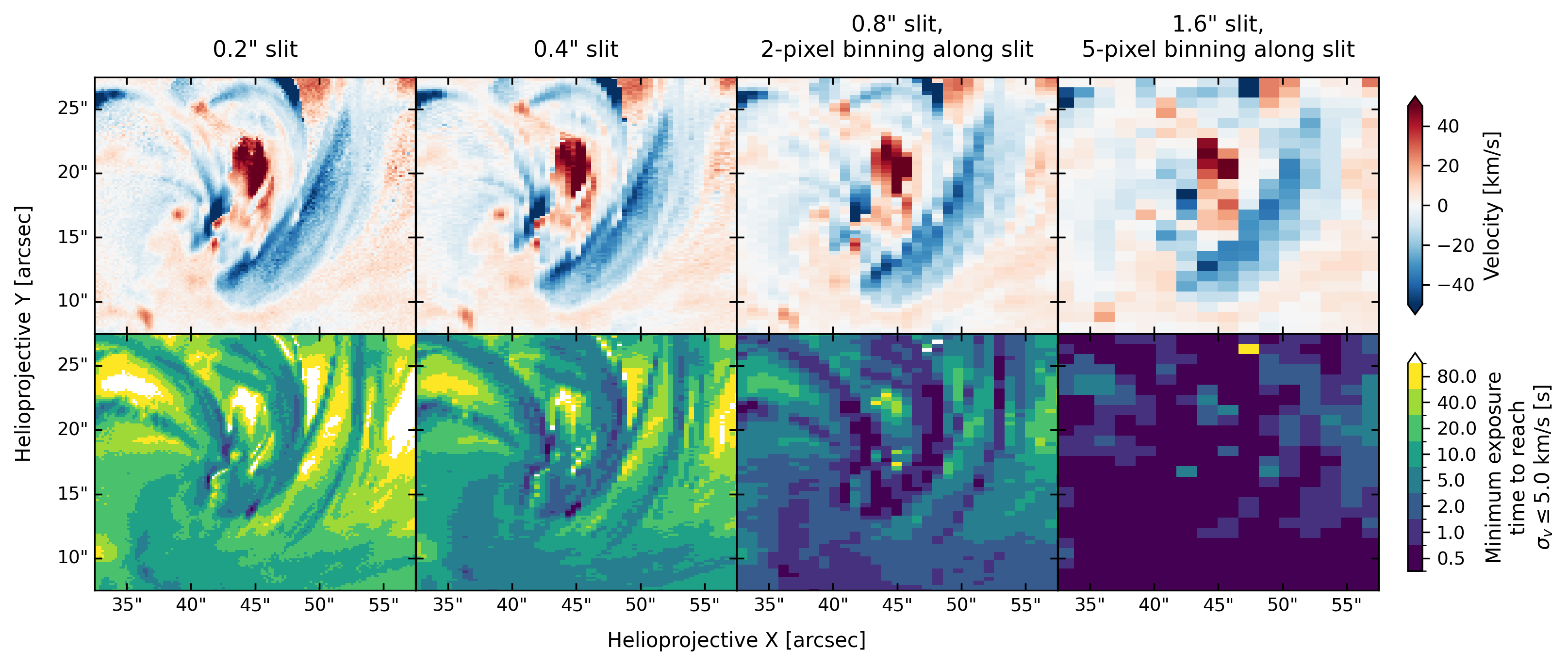}
    \caption{Fe~XII~195.119~\AA\ Doppler velocity measurements of pre-flare coronal plasma by SOLAR-C/EUVST for various slit widths \psfupdate{and along-slit binning} (top) and the minimum exposure time required to reach a measurement precision (and accuracy; see Section~\ref{sec:measurement_accuracy}) \psfupdate{of 5~km/s (bottom)}, calculated from the 1$\sigma$ centroid variation in DN/pix fitted Gaussians of 500 Monte Carlo iterations ($\sim$4\% statistical precision). Doppler velocity measurements in the top row use the longest exposure results to represent spatial resolution variation alone.\\
    Alt text: 8-panel figure arranged in two-by-four grid, with top row showing velocity maps, and the bottom row showing maps of required exposure times, for various slit widths shown in the columns.}
    \label{fig:multislit_velocity_exposure_times}
\end{figure*}

In Figure~\ref{fig:multislit_velocity_exposure_times} we show the exposure times required to reach a Doppler velocity precision (and accuracy) of 
\psfupdate{5~km/s}, for the Fe~XII~195.119~\AA\ emission line and for the slit widths available to SOLAR-C/EUVST, in the pre-flare region. We also show the Doppler velocity measurements with these slit widths (top row). 
The velocities of interest measured in this region are relatively high and so we consider a 5~km/s measurement accuracy more appropriate in this context, particularly given the priority for faster raster scanning of such dynamic plasma. With this in mind, we see that exposure times of around \psfupdate{20~seconds} per slit position are sufficient for the 0.2~arcsec slit
. \psfupdate{These exposure time requirements approximately halve with each doubling of the slit width, and halve again when pixel binning is doubled. We find that our fastest observing rate of 0.5~seconds can be achieved with the 0.8~arcsec slit and 2-pixel binning in the core of the pre-flare region.}

\begin{figure}
  \centering
  \includegraphics[width=\linewidth]{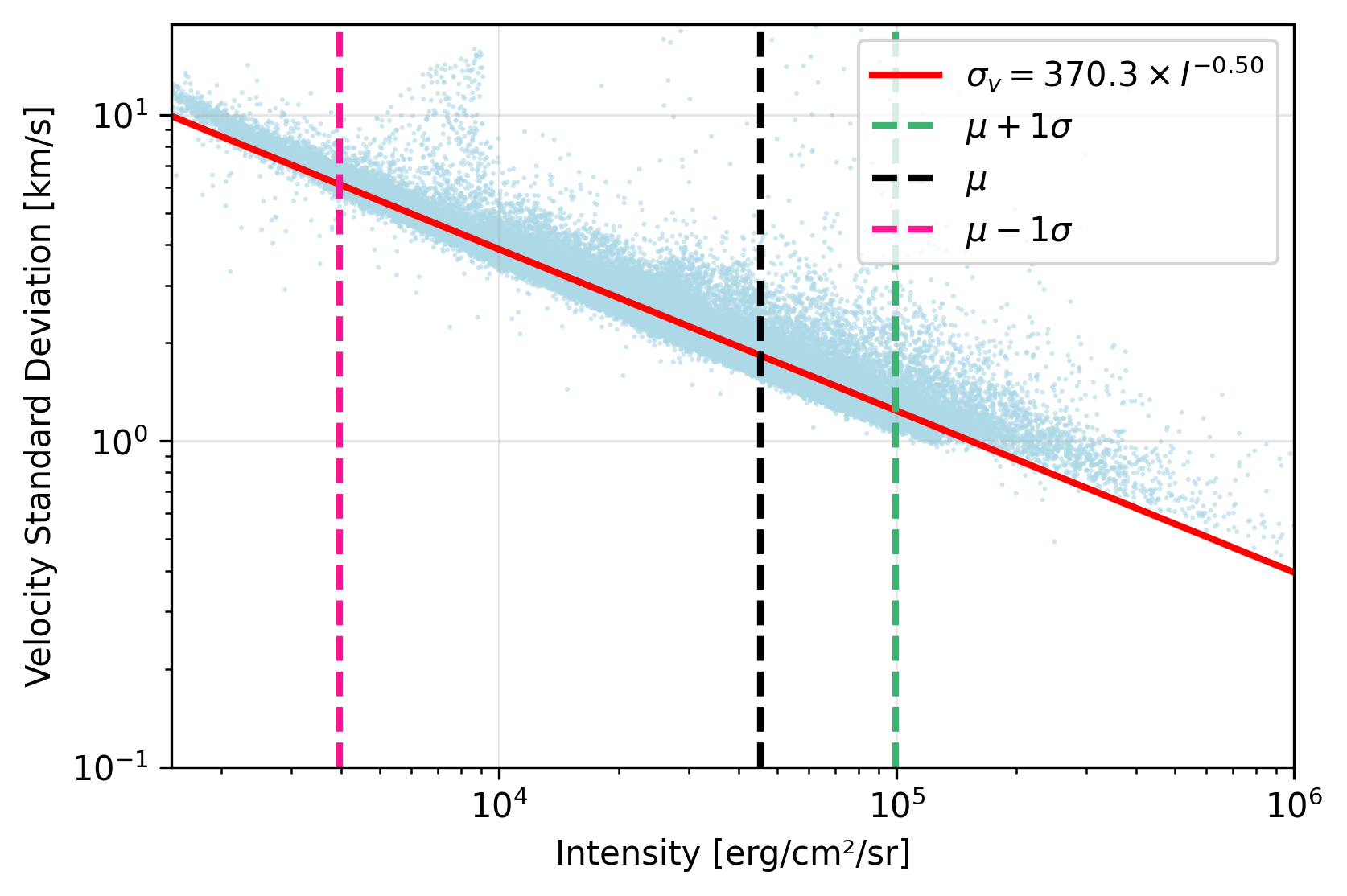}
  \caption{Scatter plot showing the relationship between Fe~XII~195.119~\AA\ intensity and Doppler velocity measurement precision (and accuracy; see Section~\ref{sec:measurement_accuracy}) for a 40-second exposure with SOLAR-C/EUVST-SW using the \psfupdate{0.4~arcsec} slit. Each point represents a single pixel from the simulated observation. \psfupdate{500} Monte Carlo iterations were used ($\sim$\psfupdate{4}\% statistical precision). Best fit line is calculated over pixels within the shown range of velocity standard deviation.\\
  Alt text: Scatter plot with logarithmic axes showing data points with overlaid red trend line and vertical dashed reference lines.}
  \label{fig:intensity_vs_velocity_std_scatter}
\end{figure}

The relationship between the intensity of emission and Doppler velocity measurement accuracy can be seen from a comparison of Figure~\ref{fig:instrument_maps} and Figure~\ref{fig:velocity_variation} in that a lower intensity is responsible for a lower measurement accuracy. We, therefore, consider this relationship for a 40-second exposure time per slit position across the active region in Figure~\ref{fig:intensity_vs_velocity_std_scatter}. We see a power law relationship between the intensity and Doppler velocity measurement accuracy, of the form \psfupdate{$\sigma_v = 370.3 \times I^{-0.50}$ km~s$^{-1}$}, where $I$ is the line intensity in physical units and $\sigma_v$ is the Doppler velocity measurement accuracy. We 
observe a spread in $\sigma_v$ beyond the $\sim$\psfupdate{4}\% statistical precision this many iterations should give. This appears to be due to the underlying plasma distributions not being truly Gaussian and so being best fitted in multiple ways with multiple resulting Doppler velocities. We discuss this in Section~\ref{sec:discussion}.

\section{Discussion}\label{sec:discussion}

\subsection{Throughput}

For the Fe~XII~195.119~\AA\ emission line we consider here, the throughput of SOLAR-C/EUVST-SW will be approximately 4~times higher than that of Hinode/EIS, but each of its pixels will collect fewer photons due to its narrower slits ($\Delta{}x$) and much finer plate scale ($\Delta{}y$). Where Hinode/EIS and SOLAR-C/EUVST-SW have a different $T_\text{opt}$, QE, $\Delta{}x$, $\Delta{}y$ and $\Delta{}\lambda$, the difference in photons reaching the detector plane per spatial pixel (integrated in wavelength, along each row of pixels on the CCD) can be calculated using $\frac{T_{\text{opt},\text{EIS}}\,\Delta{}x_\text{EIS}\,\Delta{}y_\text{EIS}}{T_{\text{opt},\text{SW}}\,\Delta{}x_\text{SW}\,\Delta{}y_\text{SW}}$. This results in Hinode/EIS collecting approximately \psfupdate{4--5~times} as many photons as SOLAR-C/EUVST at this wavelength per emission line per spatial pixel when using their 1~arcsec and \psfupdate{0.4~arcsec} slits respectively
. When spectrally resolving each emission line on the CCD, as SOLAR-C/EUVST-SW has a higher spectral sampling, the number of photons per CCD pixel is further reduced slightly by $\frac{\Delta{}\lambda_\text{SW}}{\Delta{}\lambda_\text{EIS}}$. 

As detailed in Section~\ref{sec:compare_wth_eis}, the Fe~XII~195.119~\AA\ emission line we study here shows the narrowest improvement in throughput by SOLAR-C/EUVST over Hinode/EIS due to the latter being specifically optimised for a strong sensitivity peak at this wavelength. The other wavelengths in the SOLAR-C/EUVST-SW waveband are expected to offer well over an order of magnitude increase in throughput. We discuss our investigations to confirm the strength of the Fe~XII~195.119~\AA\ signal expected from the camera in later sections.

\subsection{Pre-flare plasma flows}

Photospheric and low chromospheric motions (where plasma $\beta\gg1$) build non-potential magnetic field structures in the corona (where plasma $\beta\ll1$), which at some point release this stored energy into the coronal plasma in a solar flare \citep[e.g.,][]{mckevitt_link_2024}. This point can be thought of as being reached by two primary pathways, the global evolution of the magnetic field reaching a preferential configuration for reconnection, or the local \lq{}triggering\rq{} of the energy release. The MHD simulation we use here relies on the emergence of a twisted magnetic bipole close to a pre-existing sunspot to trigger a solar flare, with the emergence region being the one seen in Figure~\ref{fig:synthetic_intensity_map} (see \citealt{cheung_comprehensive_2018} for details).

\begin{figure}
    \centering
    \includegraphics[width=\linewidth]{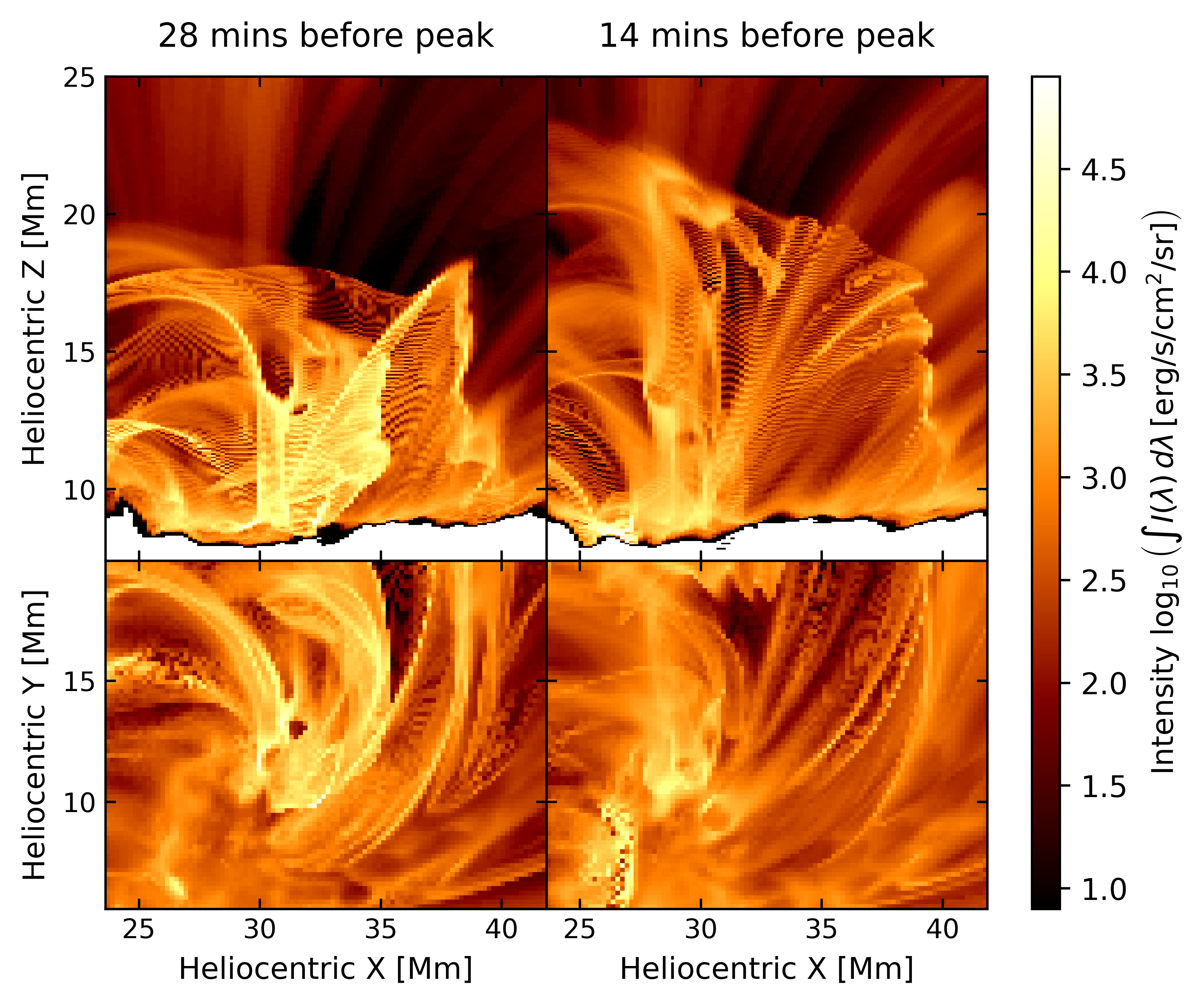}
    \caption{Synthetic Fe~XII~195.119~\AA{} emission of pre-flare region 28 minutes before the GOES peak (left) and 14 minutes before GOES peak (right) integrated along the Y axis (top) and along the Z axis (bottom). Where integrating along the Y axis, only the pre-flare plasma shown within the upper-right box in Figure~\ref{fig:instrument_maps} was used, to better isolate the pre-flare structure.\\
    Alt text: Four-panel figure arranged in two-by-two grid, with top row showing side-view images and bottom row showing overhead-view images at two time points.}
    \label{fig:side_on_intensity_map_series}
\end{figure}

During such emergence-led flare triggering, low-altitude high-velocity flows are expected to be generated due to magnetic reconnection between the pre-existing magnetic field structure and the one generated by the newly emerged flux. A good discussion of such local triggering field interacting with pre-existing coronal loops using simulations and observations can be found in \cite{kusano_magnetic_2012} and \cite{toriumi_magnetic_2013}. Above and below the reconnection point, one can expect strong upflows and downflows respectively which, when viewed from above, could be seen as adjacent plasmoids of strong opposite Doppler flows, such as those we see from SOLAR-C/EUVST-SW in Figure~\ref{fig:instrument_velocity_flare}. We consider a later time in the simulated MHD atmosphere to further analyse this. 

Figure~\ref{fig:side_on_intensity_map_series} shows the synthesised Fe~XII~195.119~\AA{} (logT$\sim$6.2) plasma from the pre-flare part of the atmosphere as viewed from above and also from the side. Only the plasma from the pre-flare region shown in Figure~\ref{fig:instrument_maps} was considered to isolate line-of-sight plasma to only that concerning these coronal structures. We consider this region at the time of all our simulated observations 28~minutes before the peak in GOES emission, and again at 14~minutes before the peak in GOES emission. We show the simulated atmosphere at full simulation resolution. We see that the simulations reveal the evolution of low-altitude fine-scale structures that are likely directly related to magnetic reconnection that triggers the eruption. Therefore, observing the dynamics of these morphologies and velocity structures with high spatio-temporal resolution is crucial for understanding the driving mechanisms of flares.

As we see the adjacent oppositely Doppler-shifted plasmoids expected from emerging flux-triggered reconnection (Figure~\ref{fig:instrument_velocity_maps_flare}), we can be confident SOLAR-C/EUVST will be capable of observing such triggering reconnection thought to be one mechanism responsible for the explosive release of energy from the solar corona. We see here clearly the effect of the Hinode/EIS PSF. The spatial extent of the plasmoids means the $\sim$1~arcsec ideal spatial resolution of Hinode/EIS would be just sufficient to resolve them, but that the $\sim$3~\psfupdate{arcsec}~FWHM PSF reduces this spatial resolution in reality along the slit direction.

Upflowing plasma is seen by Hinode/EIS as enhanced in regions before a flare occurs \citep[e.g.,][]{imada_coronal_2014}, possibly related to early-onset slow reconnection \citep{woods_observations_2017}. We see in Figure~\ref{fig:instrument_velocity_line_profile_flare} that SOLAR-C/EUVST-SW will measure much higher velocities than Hinode/EIS, meaning it may be possible to see such a pre-flare signature earlier than is currently possible with Hinode/EIS.

These flows encompass one of several spectroscopic signatures known to precede solar flares (see Section~\ref{sec:introduction} and \citealt{harra_coronal_2023}). Another key one, the excess (non-thermal) broadening of spectral lines, has also been seen using Hinode/EIS \citep[e.g.,][]{harra_location_2013}. There is a known link between upflowing plasma and non-thermal broadening as seen by Hinode/EIS referenced by \cite{imada_coronal_2014} in relation to pre-flare upflowing plasma, and something widely accepted to be due to unresolved high-speed upflows \citep{doschek_flows_2008, del_zanna_solar_2018}. \cite{doschek_dynamics_2012} identified the need to build an instrument capable of resolving such flows and remove this degree of ambiguity from the many mechanisms thought to contribute to non-thermal broadening. Once the expected PSF of SOLAR-C/EUVST is better characterised through ground measurements of key components, we will investigate a similar comparison to Figure~\ref{fig:instrument_velocity_maps_flare} but considering excess broadening. We can, however, expect smaller excess broadening given SOLAR-C/EUVST will resolve plasma on smaller scales. For example, the Hinode/EIS pixels on the boundary between the upflowing and downflowing plasmoids seen in Figure~\ref{fig:instrument_velocity_maps_flare} draw in plasma from either side of the boundary due to the PSF. We see this here as a suppression of measured Doppler velocities, but this would also increase spectral line broadening. This is promising for our ability to better localise and understand the physical drivers of excess broadening increases before solar flares with SOLAR-C/EUVST.

Our work here also suggests that pre-flare upflows are not limited to compact regions but can extend over relatively broad areas of active regions. To resolve their spatial structure and temporal evolution in detail, it is essential to observe with both wide field-of-view and short time cadence. Here, coordination with the Multi-slit Solar Explorer \citep[MUSE;][]{de_pontieu_multi-slit_2020} will be useful. While SOLAR-C/EUVST will provide detailed spectroscopic diagnostics over a wide temperature range, MUSE offers high-cadence, multi-slit observations optimised for capturing fast-evolving structures in the corona. The synergy between the two instruments will allow us to track the build-up and release of energy across multiple layers of the solar atmosphere across wide regions, and to examine the physical conditions leading up to flare onset with a level of coverage and diagnostic accuracy not previously achievable.

\subsection{Active region loops}

Previous work has shown the loops seen by Hinode/EIS to have narrow thermal distributions, consistent with them being fully or nearly fully resolved \citep{brooks_solar_2012}. Current nanoflare theory relies on much larger numbers of fundamental strands braiding and heating impulsively \citep[e.g.,][]{klimchuk_solving_2006}, rather than a handful of larger flux tubes as observations of \cite{aschwanden_elementary_2005} and \cite{brooks_solar_2012} using the Transition Region and Coronal Explorer \citep[TRACE;][]{handy_transition_1999} and Hinode/EIS respectively suggest.

The MHD-simulated atmosphere we consider here produces a small number of quasi-parallel flux tubes making up loops, consistent with the work of \cite{aschwanden_elementary_2005} and \cite{brooks_solar_2012}. However, this atmosphere is limited by simulation resolution, and dedicated simulations of nanoflare heating are better suited to specifically address the spectroscopic signatures of such a mechanism (see e.g. \cite{patsourakos_nonthermal_2006}). Nevertheless, while the MHD simulation we consider does not let us comment on the potential substructure of these flux tubes, it does allow us to determine that if active region loops are structured on scales just marginally beyond the current limits of our instrumentation, SOLAR-C/EUVST-SW will be capable of resolving such substructure. While dedicated simulations of the EUVST-LW channel's response to this atmosphere are reserved for future work, it uses the same slit, has a near identical plate scale, and comparable PSF, meaning we can expect SOLAR-C/EUVST as a whole instrument to sufficiently resolve the substructure of active region loops at a wide temperature range.

This substructure resolved by SOLAR-C/EUVST in the active region loop also displays some interesting flow features. The southern-most strand in the loop appears to show upflowing plasma to the east and downflowing plasma to the west (see lower most annotated dashed line in bottom panel of Figure~\ref{fig:instrument_velocity_maps_loop}), where the northern quasi-parallel strand appear continually upflowing (upper dashed line in bottom panel of Figure~\ref{fig:instrument_velocity_maps_loop}). Siphon flows would logically be present in flux tubes which posses different gas pressures at either end due to being rooted in different magnetic field strengths \citep{meyer_model_1968}, and while observational evidence is limited (see \citealt{guglielmino_high-resolution_2011}) they are present in MHD-simulated loops \citep[e.g.,][]{peter_asymmetries_2010}. For one strand of two which make up an active region loop to display siphon flows while both strands are rooted in almost identical magnetic (and so also gas pressure) environments, is curious. The intensity of the southern strand is seen to decrease towards the west at the point where the plasma is measured as switching from upflowing to downflowing, and so the downflow measurement could be influenced by lower intensity plasma around the strand which we see in Figure~\ref{fig:instrument_velocity_maps_loop} is downflowing. Further investigation is required to pinpoint the precise nature of this measured velocity reversal, but our synthetic observations allow us to determine that SOLAR-C/EUVST-SW will be capable of resolving siphon flows in loops and their sub-structure, provided such substructure is comparable to that in this MHD simulation and as described by \cite{aschwanden_elementary_2005} and \cite{brooks_solar_2012}.

At loop footpoints, \cite{hara_coronal_2008} found enhanced non-thermal line widths and blue-wing asymmetries seen by Hinode/EIS that are consistent with superposed unresolved high-speed upflows. With the spatial resolution improvements we have demonstrated, we expect SOLAR-C/EUVST to separate these into more distinct Doppler components and to higher velocities. 






\subsection{Measurement accuracy}

As charge builds up on the short wavelength CCD during a SOLAR-C/EUVST-SW exposure, the signal to noise ratio increases as the strength of the signal increasingly outweighs the various noise effects we describe in Section~\ref{sec:method}. As a slit spectrograph, separate exposures are required at each X position of mapping the solar atmosphere, making large active regions time consuming to fully map and putting importance on understanding the minimum time required to have a sufficient signal to noise ratio to meet a high Doppler velocity measurement accuracy.

The precision of Hinode/EIS when designed was to be better than 5~km/s for exposure times of between 10~s and 100~s for active regions \citep{culhane_euv_2007}, and measurements indicate it to be precise to around 1--2~km/s \citep[e.g.,][]{mariska_doppler-shift_2010}. However, because of the movement of spectral lines across the detector due to thermal deformation of the spacecraft, an additional pipeline step is required to correct for such a shift, meaning Doppler velocity measurements are only accurate to $\sim$4.4~km/s \citep{kamio_modeling_2010}. 
We note that as discussed in Section~\ref{sec:measurement_accuracy} we take our calculated precision values as measurement accuracy as well. Figure~\ref{fig:velocity_variation} shows that, for a 40~s exposure per slit position, the Doppler velocity accuracy of SOLAR-C/EUVST-SW is better than that of Hinode/EIS for almost all of the active region, and reaches as low as 1~km/s at loop footpoints, in loops, and in the pre-flare region. 


Figure~\ref{fig:exposure_time_requirement} shows that in this MHD simulation snapshot 30~minutes before the flare peak, the intensity of plasma emission is already sufficient to allow exposure times of \psfupdate{down to 2}~seconds to meet 2~km/s measurement accuracy. To meet this accuracy in the quiescent active region plasma, exposures of around \psfupdate{10~s to 20~s} per slit position are required. For footpoints separated by around 50~arcsec as those in this active region, using the \psfupdate{0.4~arcsec} slit would require \psfupdate{125} raster scanning steps, with a total scanning time of between 20 and 40~minutes. SOLAR-C/EUVST will have a pointing accuracy of around 20~arcsec and a pointing stability on the order of 0.2~arcsec (3$\sigma$) at these scanning times, and so including margins on targetting the raster scan, we can expect to be able to take such measurements of an active region loop system at this temperature with the Fe~XII~195.119~\AA\ emission line in times between 30~minutes and an hour. This can be reduced by using any of the larger slits available to SOLAR-C/EUVST, by rastering a smaller part of the active region, or any of the other methods discussed in Section~\ref{sec:results_spatial}.

The relationship between the intensity of a spatial pixel and the Doppler velocity accuracy we meet there for a 40-second exposure, shown in Figure~\ref{fig:intensity_vs_velocity_std_scatter}, shows a clear power law relationship. If governed entirely by shot noise, due to its Poissonian nature we would expect the exponent of such a relationship to be $-0.5$\psfupdate{, which we obtain here.} For the range of pixel values we show for this 40-second exposure, covering the $\mu+\sigma$, $\mu$ and $\mu-\sigma$ intensities, shot noise is the dominant source of noise and the various sources of detector noise play a comparably small role. 
\psfupdate{In shorter exposures or in observations of lower-intensity plasma}, we expect a larger contribution of the overall noise to come from detector effects.

The evolution of pre-flare plasma such as we showed in Figure~\ref{fig:instrument_velocity_flare} is highly dynamic, and an emphasis in such observations is placed on the rapid scanning of such a region to capture as much of this evolution as possible. While sacrificing spatial resolution, wider slits offer compound benefits to raster cadence. When the slit width is doubled so is the number of photons collected (see Equation~\ref{equ:photons_subtended_angle}). In shot-noise governed exposures (such as those of our instrument in this case), this results in the signal-to-noise ratio doubling and so the exposure time per slit position approximately halving. When the slit width is doubled, the required number of scanning steps is also halved, meaning in the case of our instrument and this pre-flare plasma regime, doubling the slit width approximately quarters the required time to perform a raster scan. \psfupdate{Using Figure~\ref{fig:multislit_velocity_exposure_times} we can determine} that the pre-flare region could be scanned with a cadence of \psfupdate{$\sim$40~minutes} with the 0.2~arcsec slit, \psfupdate{and that} this reduces to \psfupdate{$\sim$10~minutes} with the 0.4~arcsec slit, 
\psfupdate{and less than 1~minute with the 0.8 and 1.6~arcsec slits (due also to the binning we consider).} 
We also see the 0.4~arcsec slit observations resolve practically all the detail of the 0.2~arcsec slit in the case of this plasma, making this slit a compelling choice when fast cadences are required with pre-flare intensities. In that case, SOLAR-C/EUVST would capture the atmospheric evolution we show in Figure~\ref{fig:side_on_intensity_map_series} in \psfupdate{2 to 3} separate rasters. A full analysis of the performance of the instrument in capturing pre-flare and flare dynamics is reserved for future work, with high-cadence MHD simulation snapshots needed which were not available for this study. We also note the advantage of the analysis we show in Figure~\ref{fig:multislit_velocity_exposure_times}. The various contributing noise regimes will vary in significance during the evolution of plasma and its intensity, making approximating the impact of widening the slit on signal-to-noise ratio complex, particularly outside the shot-noise limited regimes. The analysis we present takes this complexity into account.

The power law function we derived effectively encodes all the detailed technical aspects of the EUVST telescope and SW detector into one simple relationship. We note though that at the point the data pipeline is built for users it is likely that dark current subtraction will be done, as is done in the Hinode/EIS pipeline, and so such a power law will be offset.
Figure~\ref{fig:instrument_key_pixel_spectra} shows the measured spectra sitting on a pedestal of around \psfupdate{25~DN/pix}. This is a combination of the build up of dark current in the CCD pixels and the background lines we synthesised (Figure~\ref{fig:synthetic_spectra}). As the dark current will build up the same in each of the detector pixels equally, any small difference between the background level seen in the $\mu-1\sigma$ and $\mu+1\sigma$ spatial pixels is due to the greater emission from the background emission lines in the higher intensity pixels (as seen in Figure~\ref{fig:synthetic_spectra}). We plan to release similar $\sigma_v=f\left(I,t\right)$ functions (being also a function of exposure time $t$) for the main emission lines of SOLAR-C/EUVST, including for the long wavelength channel with its different noise profile, once the instrument is calibrated. We expect users of SOLAR-C/EUVST data to use such functions in observations to quantify the uncertainties in their velocity measurements.

Of interesting note in Figure~\ref{fig:intensity_vs_velocity_std_scatter} is that 
we observe for each intensity a spread in the velocity standard deviations. We attribute this spread above the statistical noise of the method to the spectral distributions not being truly Gaussian on average and so Gaussian fits not accurately capturing the centroid location. Although we use a Maxwellian distribution of velocities in our emission line synthesis (Equation~\ref{equ:synthetic_gaussian_line_profile}), as our spectrograph integrates through all the plasma along the line of sight, this plasma exists with various Doppler shifts (seen in Figure~\ref{fig:synthetic_dems}) so we will rarely find a true Gaussian distribution of line-of-sight plasma.

\section{Conclusions}

In this paper we have presented the results from a new forward modelling code which demonstrate the performance of the short wavelength camera on board SOLAR-C/EUVST. The code includes all relevant instrumental effects and noise sources, and is a complete pipeline which takes intensity in a simulated solar atmosphere and converts it to DN/pixel as generated by the detector's readout electronics and transmitted to Earth.

We used this code to simulate observations of a pre-flare active region from SOLAR-C/EUVST-SW, and consider the ability of the instrument to resolve, in space, plasma motion and structure in the active region. As the spatial resolution and PSF is expected to be near-identical in the LW channel, we draw conclusions about the whole EUVST spectrograph.


The MURaM atmosphere we used contains several distinct active region loops separated by several hundred kilometres. Our simulations of SOLAR-C/EUVST suggest the instrument will resolve these loops on such spatial scales, which some literature suggests is the fundamental size of coronal structures. Future observations of such structures is essential to better understand coronal heating mechanisms.

We also find SOLAR-C/EUVST able to detect the signatures of low-altitude local flare triggering mechanisms, in this case the reconnection of a pre-existing magnetic field structure with one generated by newly emerged twisted flux 30-minutes before the flare peak.


\begin{ack}
  We thank the reviewer for their comments which improved our manuscript. We thank Matthias Rempel and Paola Testa for discussing MHD simulation timesteps. Synthesis and simulations used Austrian Super Computing (ASC) infrastructure. \psfupdate{This research used version 0.6.2 \citep{will_barnes_fiasco_2025} of the fiasco open source software package. GitHub Copilot was used to assist with code review.} We thank the European Space Agency (ESA) for funding the development of the SW camera for SOLAR-C/EUVST, and JAXA and all other relevant agencies for their contributions to the mission. We thank the engineers at PMOD in Davos for a useful discussion of contamination effects, and Marie Dominique at ROB for their assistance with optical filter information. We thank NRL (U.S.) and ISAS/JAXA (Japan) for hosting J.M. to complete this work. We thank Mark Cheung and Lucie Green for their helpful comments on our work.
\end{ack}

\section*{Funding}
  J.M. was supported by STFC PhD Studentship number ST/X508858/1. S.M. and D.B. acknowledge support from UKSA grant No. UKRI920. S.M. was also supported by ESA Contract No. 4000141160/23/NL/IB. The work of D.H.B. was performed under contract to the Naval Research Laboratory and was funded by the NASA Hinode program. T.S. was partially supported by the grant of OML Project by the National Institutes of Natural Sciences (NINS program No. OML032402). A.T. was supported by JSPS KAKENHI Grant Nos. JP23KJ2151 (PI: A.T.) and JP21K13971 (PI: A.T.). S.T. was supported by JSPS KAKENHI Grant Nos. JP21H04492 (PI: K. Kusano), JP25K01041 (PI: K. Namekata). T.M.D.P. was supported by the Research Council of Norway through its Centres of Excellence scheme, project number 262622. Hinode is a Japanese mission developed and launched by ISAS/JAXA, with domestic partner NAOJ and international partners NASA and STFC/UKSA (UK). It is operated in cooperation with ESA and NSC (Norway).

\section*{Data availability} 
  The \texttt{ECLIPSE} forward modelling code is available online, alongside tutorials and a Python API: \url{https://github.com/jamesmckevitt/eclipse}. We use version \psfupdate{0.6.1.4} for this paper, found here: \psfupdate{\url{https://doi.org/10.5281/zenodo.19662222}}.

\appendix

\section*{Comparison with isothermal-monovelocity approximation}

When synthesising spectral line intensities from MHD simulations, care needs to be taken that such a synthesis produces realistic spectra. The 3D MHD simulation we consider here provides single values of temperature, density, and velocity for each computational voxel, yet real plasma exists as a quasi-continuous distribution. We compare two approaches for handling this discretisation and explain our choice of methodology.

The most straightforward approach treats each simulation voxel as containing plasma at exactly the temperature and velocity specified by the simulation. For each voxel, the emission is calculated assuming isothermal plasma moving at a single velocity, then individual contributions are summed along the line of sight:

\begin{multline}\label{equ:isothermal_monovelocity}
    I_{ij}(\lambda) = \sum_{z} \frac{{{N_e}_z}^2\,G_{ij}(T_z,{N_e}_z)}{4\pi} \times \phi(\lambda,T_z,v_z) \times \Delta h \\ [\text{erg s}^{-1}\text{ cm}^{-2}\text{ sr}^{-1}\text{ cm}^{-1}],
\end{multline}

\noindent{}where $\phi(\lambda,T,v)$ is the same as defined in Equation~\ref{equ:synthetic_gaussian_line_profile}.

This method is more computationally simple and preserves the exact values computed by the simulation. Each voxel contributes a Gaussian centered at its Doppler-shifted wavelength with width determined by its temperature. For relatively uniform plasma conditions where the simulation sufficiently samples the plasma, this works well and provides a direct representation of the simulation data.

Our method recognises that simulation voxels represent volume-averaged quantities rather than delta functions in parameter space. We bin the voxels into a two-dimensional histogram in temperature-velocity space, creating the distribution $\Xi(T,v)$ as described in Section~\ref{sec:2d_dem_equation}. This transforms the discrete voxel data into a smoother representation before calculating line profiles.

The key difference is that rather than treating the simulation as providing exact point samples of plasma properties, we interpret it as sampling an underlying continuous distribution. The binning process effectively performs a kernel density estimation, acknowledging that real plasma transitions smoothly between states rather than jumping discontinuously. For quiescent active regions with modest velocity gradients, both methods produce similar results. The thermal broadening naturally smooths the contributions from adjacent voxels, and the resulting line profiles are comparable. However, in dynamic regions, particularly the pre-flare atmosphere we study, the methods diverge. The emission measure approach can introduce smoothing that obscures fine structure if the binning is too coarse, and so we chose our bin widths to accurately capture the plasma from this MHD simulation.

\makeatletter
\global\@in@appendixfalse
\makeatother
\bibliographystyle{aa}
\bibliography{references}

\end{document}